\documentclass[10pt]{article}
\usepackage{geometry}
\title{Massless scalar free Field in 1+1 dimensions I:\\ Weyl algebras Products and Superselection Sectors}
\author{Fabio Ciolli \\
 \small{Dipartimento di Matematica, Universit\`a di Roma ``Tor Vergata'' }\\
    \small{Via della Ricerca Scientifica I-00133, Roma,  Italy}  \\
          \small{\texttt{ciolli@mat.uniroma2.it}}}

\usepackage{latexsym, amssymb, amsmath}
\def\1{{\mathbf 1}}
\def\0{{\mathbf 0}}
\def\csv{{C^*(V, \s_V)}}
\def\csvn{{C^*(V_0, \s_{V_0})}}
\def\csh{{C^*(H, \s_H)}}
\def\csl{{C^*(L, \s_L)}}
\def\csco{{C^*(C)}}
\def\csn{{C^*(N)}}
\def\dC{{\widehat{C}}}
\def\csa{{C^*(V_a, \s_a)}}
\def\csb{{C^*(V_b, \s_b)}}
\def\csc{{C^*(V_c, \s_c)}}
\def\csq{{C^*(V_q, \s_q)}}
\def\cse{{C^*(V_e, \s_e)}}
\def\csf{{C^*(V_f, \s_f)}}
\def\Zb{{\cZ_b}}
\def\catV{{\mathbf{Symp}}}
\def\catW{{\mathbf{Weyl}}}

\def\OL{{\O_L}}
\def\dG{{\widehat{\cG}}}
\def\tmax{{\otimes_{\rm max}}}
\def\tmin{{\otimes_{\rm min}}}
\def\dS{{\partial \cS}}
\def\duS{{\partial^{-1} \cS}}
\def\duzS{{\partial^{-1}_0 \cS}}
\def\duqS{{\partial^{-1}_q \cS}}%
\def\dt{{\partial t}}
\def\T{{\partial t\oplus t}}
\def\fzu{{f_0\oplus f_1 }}
\def\Fl{{F_c \oplus F_n }}
\def\Fm{{F_r\oplus  F_q }}
\def\Ft{{\fzdt\oplus\fut}}
\def\Fin{{F_\infty}}
\def\fzdt{{f_{0\dt}}}
\def\fut{{f_{1t}}}
\def\gzu{{g_0 \oplus g_1 }}
\def\psii{{\psi_\infty}}
\def\timeinfty{{\underset{\infty}{\bowtie}}}
\def\wL{{\cW_L}}
\def\cA{{\cal A}}
\def\cB{{\cal B}}
\def\cC{{\cal C}}

\def\cE{{\cal E}}
\def\cF{{\cal F}}
\def\cG{{\cal G}}
\def\cH{{\cal H}}
\def\cI{{\cal I}}

\def\cK{{\cal K}}
\def\cL{{\cal L}}

\def\cN{{\cal N}}
\def\cO{{\cal O}}

\def\cP{{\cal P}}
\def\cQ{{\cal Q}}
\def\cR{{\cal R}}
\def\cS{{\cal S}}
\def\cT{{\cal T}}
\def\cU{{\cal U}}

\def\cW{{\cal W}}

\def\cZ{{\cal Z}}

\def\bC{{\mathbb C}}

\def\bR{{\mathbb R}}
\def\bT{{\mathbb T}}

\def\bbC{{\mathbf{C}}}%
\def\bbN{{\mathbf{N}}}%
\def\bbU{{\mathbf{U}}}%
\def\bbW{{\mathbf{W}}}%
\def\a{\alpha}
\def\b{\beta}
        											\def\G{\Gamma}
\def\d{\delta}        											\def\D{\Delta}

\def\z{\zeta}       
\def\e{\eta} 
\def\th{\theta}          \def\Th{\Theta}
\def\i{\iota}

\def\l{\lambda}       \def\L{\Lambda}
\def\m{\mu}

\def\x{\xi}
\def\p{\pi} 
\def\r{\rho}
\def\s{\sigma}

\def\f{\varphi}       

\def\c{\chi}
\def\o{\omega}        \def\O{\Omega}

\def\fA{{\mathfrak A}}

\def\fF{{\mathfrak F}}

\def\fH{{\mathfrak H}}

\def\fN{{\mathfrak N}}

\def\fZ{{\mathfrak Z}}

\def\to{\rightarrow}

\def\ov{\overline}
\def\supp{{\mathrm{supp\,}}}
\def\loc{{\mathrm{loc\,}}}
\def\ad{{\mathrm{ad\,}}}
\def\aut{{\mathrm{Aut\,}}}
\def\inn{{\mathrm{Inn\,}}}
\def\repb{{\mathrm{Rep^\bot\,}}}
\def\open{{\mathrm{Open\,}}}
\def\sub{{\mathrm{Sub\,}}}
\newcommand{\abs}[1]{|#1|}
\newcommand{\norm}[1]{\| #1 \|}
\def\rest{{\upharpoonright}}

\def\nbot{\vbox {\hbox to 10pt {$\not$\hskip -1.3pt $\bot$\hfil}}}
\newtheorem{thm}{Theorem}[section]

\newtheorem{dfn}[thm]{Definition}
\newtheorem{lemma}[thm]{Lemma}
\newtheorem{prop}[thm]{Proposition}
\newtheorem{cor}[thm]{Corollary}

\newtheorem{rem}[thm]{Remark}
\newtheorem{axi}[thm]{Axiom}
\newcommand{\baxi}{\begin{axi}}
\newcommand{\eaxi}{\end{axi}}
\newcommand{\bassum}{\begin{assum}}
\newcommand{\eassum}{\end{assum}}
\newcommand{\bdfn}{\begin{dfn}}
\newcommand{\edfn}{\end{dfn}}
\newcommand{\blemma}{\begin{lemma}}
\newcommand{\elemma}{\end{lemma}}
\newcommand{\bprop}{\begin{prop}}
\newcommand{\eprop}{\end{prop}}
\newcommand{\bthm}{\begin{thm}}
\newcommand{\ethm}{\end{thm}}
\newcommand{\bcor}{\begin{cor}}
\newcommand{\ecor}{\end{cor}}
\newcommand{\prf}{\noindent{\it Proof. }}
\newcommand{\qed}{\ \hfill $\square$ \\\\}
\newcommand{\bconj}{\begin{conjecture}}
\newcommand{\econj}{\end{conjecture}}
\newcommand{\brem}{\begin{rem}}
\newcommand{\erem}{\end{rem}}
\newcommand{\beq}{\begin{equation}}
\newcommand{\eeq}{\end{equation}}
\newcommand{\beqn}{\begin{equation*}}
\newcommand{\eeqn}{\end{equation*}}
\newcommand{\barr}{\begin{array}}
\newcommand{\earr}{\end{array}}
\newcommand{\beqa}{\begin{eqnarray}}
\newcommand{\eeqa}{\end{eqnarray}}
\newcommand{\beqan}{\begin{eqnarray*}}
\newcommand{\eeqan}{\end{eqnarray*}}

\newcommand{\bdes}{\begin{description}}
\newcommand{\edes}{\end{description}}
\newcommand{\bitem}{\begin{itemize}}
\newcommand{\eitem}{\end{itemize}}
\numberwithin{equation}{section}

\begin{document}
\maketitle\noindent
\begin{abstract}
This is the first of two papers on the superselection sectors of the conformal model in the title, in a time zero formulation. A classification of the sectors of the net of observables as restrictions of solitonic (twisted) and non solitonic (untwisted) sector automorphisms of proper extensions of the observable net is given. All of them are implemented by the elements of a field net in a non-regular vacuum representation and the existence of a global compact Abelian gauge group is proved. 
A non trivial center in the fixed-point net of this gauge group  appears, but in a unphysical representation and reducing to the identity in the physical one.
The completeness of the described superselection structure, to which the second paper is devoted, is shown in terms of Roberts' net cohomology. 
Some general features of physical field models defined by twisted cross products of Weyl algebras in non-regular representations are also presented.
\end{abstract} 

\bigskip
\noindent
\emph{Keywords}: Weyl algebras; Massless scalar free field; Superselection sectors; Conformal models; Solitonic sectors; Twisted crossed products; Non-regular representations.

\bigskip
\noindent 
Mathematics Subject Classification 2000: 81T05, 81T10, 81T40, 46L60, 46N50

\tableofcontents
\section{Introduction}
The general theory of superselection sectors in low dimensional Quantum Field Theory is still lacking so the study of special models is of great interest. Relevant progress has been achieved in the past years using Algebraic Quantum Field Theory for various classes of models such as loop groups, orbifold models and coset models, see \cite{EK} and \cite{sch06} also for a historical review. 

In this approach a complete classification  of \emph{rational models} (i.e. with a finite number of sectors) with Virasoro central charge $c <1$, see \cite{klm01,kl04}, and for the local extensions of \emph{compact type} of the Virasoro net on the circle at $c=1$ has been attained, see \cite{car04} for details. 
In purely massive theories the triviality of sectors has been proved in \cite{mug98}.
Two notable features of the theory are the presence of topological sectors of solitonic origin, see \cite{klx04,mug04}, and the following  dichotomy between the rational and non rational case established in \cite{lx04}: if in a model all irreducible sectors have conjugates then, it is  either rational or has uncountably many different irreducible sectors. 

\smallskip
In this paper we deal with a non rational $c=1$ model, the massless scalar free field, also called  \emph{the Streater and Wilde model} after its first formulation using the ideas of local nets in \cite{sw70}. 
Our main goal is to understand better  the interplay between chiral and time zero formulation of the superselection structure in 1+1-dimensional theories, the nature of solitonic and non solitonic sectors, the relation between DHR sectors and the presence of a quantum global internal symmetry, usually reducing to a global compact gauge group. 

The results obtained in this paper, and in its sequel \cite{fabio4}, are largely adaptable to other theories based on Weyl algebras and give a strong further application of Roberts' net cohomology for discussing superselection sectors in general spacetime context, see \cite{rob00} for reference.

\smallskip
The observables of the Streater and Wilde model are obtained from the quantized derivatives of the classical fields by imposing a constraint condition on the solutions of the two dimensional wave equation. This choice avoids infrared divergences, see \cite{sw70} and references therein, and for this reason the model is also called \emph{the theory of the potential of the field}. 

The existence of a (physical, separable) Hilbert space representation for the observable net is obtained in the usual way by the Fock space second quantization procedure for Weyl algebras.  
For the same net, the existence of a continuous, i.e. uncountable, family of DHR superselection sectors is known from the original description in \cite{sw70} (see \cite{H, rob00} for general references on the Doplicher, Haag, Roberts theory of superselection sectors). They are labeled by pairs of real numbers, i.e. by the elements of $\bR_d^2$ (here the subscript $d$ means discrete topology) and realized as inner automorphisms by some left/right movers (solitons) of a field net extension. 
Other relevant features of the model were discussed in \cite{hl82}, namely the Tomita Takesaki modular structure, spacelike and timelike duality.

The usual chiral net formulation  of the Streater and Wilde model is treated in \cite{bmt88} and the above cited extension may be  classified according to \cite[Definition 3.2]{car04} as  being of compact type.

\smallskip
A relevant step in the description of Weyl algebras models was the introduction of \emph{non-regular representations}, see \cite{tn92} and \cite{ams93,ams93a}: for such a representation $\p$ of $\cW(V,\s_V)$, the Weyl algebra on the symplectic space $(V,\s_V)$, there is a subspace of elements  $v\in V$ such that, for $\l\in \bR$, the map $\l\mapsto \p(W(\l v))\subset \cB(\cH_\p)$ is not weakly continuous. 

The papers just cited pointed out the utility of such Hilbert space representations in the presence of a theory with uncountable many sectors. This  avoids the use of inner product spaces with indefinite metrics, for the (unphysical) representation of charged fields. However, the same papers do not attack the problem of a local net theory in a non-regular representation, nor the full description of its DHR sectors and associated gauge group.

\smallskip
The task of this first paper is to collect the known results on the Streater and Wilde  model and reconstruct the DHR sectors theory using a putative field net $\fF$ (that do not satisfies the locality condition) and a compact  group $\cG$ of gauge symmetries, such that the observable net $\fA$ is the fixed-point net under the action of $\cG$, restricted to the representation Hilbert space $\cH_a$ of the observables, i.e. $\fA= \fF^\cG\rest_{\cH_a}$.
In the second paper we legitimate the net $\fF$ as the \emph{complete field net} of $\fA$, and the description of the superselection structure is hence similar to the higher dimensional one of a \emph{field system with gauge symmetry}, see the celebrated \cite{dr90} for definitions and results, apart from the presence of braid symmetry instead of permutation symmetry.
\footnote{
It should be said that the superselection theory of the analogous model in $1+3$ dimensions is known to be trivial. Here a constrain condition on the symplectic space (not introduced to avoid the infrared divergences as in the 1+1 case) distinguishes the observable net $\fA$ from its Haag dual net $\fA^d\supset \fA$ and the failing of Haag duality for the observable net $\fA$, i.e. $\fA^d \neq\fA$, denotes the presence of a spontaneous breaking of the gauge group. The above mentioned triviality is due to the absence of sector (or solitonic) automorphisms for $\fA^d=\fF$, the dual net equates the field net,  and is proved by net cohomology in \cite{bdlr92}.
}    
\smallskip

To construct such a field net, together with the gauge group structure, we use the abstract tool of \emph{2-cocycle twisted crossed product of Weyl algebras}, i.e. the reinterpretation of the Weyl algebras of fields as an extension of that of the observables by the cocycle twisted action of the charge group. This \emph{simple current extension} is defined by a (generalized) 2-cocycle derived from the symplectic form of the Weyl algebras, in its observable and charge-gauge group components, already partially studied in the physical literature, see for example \cite{her98}.
\footnote{More general examples of simple current extensions, derived from loop groups,  orbifold models or vertex operator algebras may be considered. See for example \cite{kl04a}.
}  

The extension of the symplectic spaces in the model has the structure $V_a \subset V_f = V_a \oplus (N \oplus C)$ where the symplectic form splits as $\s_f = \s_a \oplus \s_{N\oplus C}$. Here, the space $C$ is the discrete Abelian charge group. The structure of the corresponding Weyl algebras is then
\beq\label{e:prima}
\cW ( V_f, \s_f ) = \cW(V_a,\s_a)\otimes \cW(N,\s_N)\rtimes_z \cU (C)\,.
\eeq
Here $\cU(C)$ denotes the Abelian group $C$ written multiplicatively. Note that the factor $\cW(N, \s_N)$ commutes with $\cW(V_a, \s_a)$, but is acted upon by the charge group through the 2-cocycle $z$, reflecting the symplectic interaction between $N$ and $C$.

\smallskip 
Such a construction, in a time zero formulation, allows one to classify the  sectors labeled by $\bR^2_d$ as restrictions of solitonic (twisted) and non solitonic (untwisted) sector automorphisms of two different simple current extensions of the net of the observables. 
Hence these sectors accord to the definitions in \cite{mug04} and \cite{klx04}, but this classification reflects a different perspective respect to the equivalent nature of the left/right solitons of the chiral formulation.  
Moreover, the time zero approach makes evident the presence of a non trivial center $\fF^\cG\cap (\fF^\cG)'$ of the fixed-point net under the action of the global compact gauge group $\cG$, weakly continuously represented on the unphysical Hilbert space.    

\smallskip
Namely, the construction by a simple current extension, gives a \emph{six-term diagram of inclusions} of (localized nets of) symplectic spaces, Weyl algebras and von Neumann algebras with the action of the charge and gauge groups. 
To introduce the test function spaces used to define the time zero symplectic spaces of the model, we denote by $\cS$ the Schwartz space of real valued rapidly decreasing functions on the real line, $\dS$ the space of functions that are derivative of functions in $\cS$ and by $\duS$ the  $C^\infty$-functions whose derivative is in $\cS$. 

Referring to the classical theory of the quantum massless field in 1+1 dimensions, if $\f$ denotes the field and $\p=\dot{\f}$ its conjugate momentum field, the currents extension corresponds to two different test function space extensions: the codimension 1 extension from $\dS$ to $\cS$, which corresponds to lifting the condition that test functions for the massless time-zero field $\f$ should vanish at zero wave number in Fourier space, and the codimension 2 extension from $\cS$ to $\duS$ which corresponds to admitting as test functions for the time-zero conjugate momentum field $\p$ both constant functions and odd functions tending to constants at infinity.    
These extensions of test function spaces, together with the extension of the corresponding symplectic form 
\[
\s_a (F, G) = \int_\bR \, (f_0 g_1 - f_1 g_0 )\, dx\,,
\]
where $F=\fzu$ and $G=\gzu$ are two different couples of test functions for field and momentum respectively, give the extension from the algebra of observables to the algebra of fields. 
We hence denote by $C$ the space of quotient classes $\tilde{f_0}(0)$ for $\f$ and by $Q$ the space of quotient classes $f_1(+\infty)-f_1 (-\infty)$  for $\p$. Together, they form a two-dimensional real space of charges $C\oplus Q$, furnished with the discrete topology, i.e. $C\cong Q\cong \bR_d$, and playing the role of the charge group of the model, denoted only by $C$ in the more generic formula (\ref{e:prima}).   

Omitting some intermediate terms, a reduced version of the cited diagram of inclusions for the (nets of) symplectic spaces and the von Neumann algebras of the observables and putative fields, is
\beqan \label{e: introduction}
V_a\,:=\,\dS\oplus \cS\,\subset \,V_b:=(\dS\oplus \cS)\oplus N\,&\subset&\,((\dS\oplus \cS)\oplus N) \oplus C\oplus Q\, \cong\,\cS \oplus \duS\,=:\,V_f\,,\\\\
\cA\,\subset\,\cB:=\cA\otimes \cZ_b\,=\,\cF^\cG\, &\subset&\,(\cA\otimes \cZ_b)\rtimes\cU(C)\rtimes\cU(Q)\,=:\,\cF\,.
\eeqan
In these diagrams,  $N\cong \bR$ denotes the space of constant test functions for $\p$, that plays the same role of $N$ in (\ref{e:prima}). The charge group acts non trivially only on the non trivial central tensor factor $\cZ_b$, the Abelian von Neumann algebra generated by the representation of $N$. 

\smallskip
A major task, that we postpone to the second paper \cite{fabio4}, is the question of the completeness of the $\bR^2_d$-labeled superselection theory. A positive answer is given by a careful choice of the index sets defining the nets and by the very effective theory of net cohomology of Roberts. Actually, we determine the sectors for a large class of models given by an extension of Weyl algebras.   
 
In the second paper the non trivial center, or relative commutant $\fA\cap \fF'$, is discussed more deeply.  This feature is considered for example in  \cite{bl03}. It will also be pointed out that it is related to the $\bR$-graded commutation rules of the non local field net $\fF$. Moreover, the relation with the superselection theory in presence of constraints, see \cite{bg04}, and  further structural properties of the nets of the Streater and Wilde model, such as duality properties with respect to different index sets and split properties, will also be considered.   

\smallskip
The structure of this paper is as follows. 

\noindent
In Section \ref{s:Weyl} we recall known and add new material on Weyl algebras, at the abstract algebraic,  C* and von Neumann algebraic level. Particularly, in Subsection \ref{ss: isom twist} a definition of the twisted cross product of Weyl algebras is given, and used to describe the observable-charge coupling in a physical model. In Subsection \ref{ss:reps} we present some useful requirements of independence of the states on the Weyl algebras of a twisted cross product, necessary for physically interesting representations, on a non separable Hilbert space.

A detailed account of non-regular representations of Weyl algebras is presented in Subsections \ref{ss:Elementary Weyl algebras}. The attention is focused on the non-regular representation of the elementary Weyl algebras on the symplectic space $L\cong \bR\oplus \bR$, presented for example in \cite{tn92}, to be used as a building block for the general twisted product case.  
\footnote{
The same algebraic characterization may be used to study the superselection structure of models presenting electromagnetic charges and interaction. In this line, the analysis of the St\"uckelberg-Kibble $QED_2$ model will be presented elsewhere. 
}

Section \ref{s:SW model} is devoted to the twisted cross product formulation of the Streater and Wilde model with initial data on the time zero line.
Fixing any charged element in the symplectic space, gives a symplectic isomorphism that exponentiates to an isomorphism of Weyl algebras. The defining representations for some intermediate and the larger putative field algebra are also introduced through this isomorphism, in an essentially unique way.

The local net theory of the model  is presented in Section \ref{s: Local theory SW model rep}, in the usual approach: the time zero observable net $\cA$ on the index set of the open bounded intervals of the time zero line is defined, so that, if $I$ is such an interval and the base of the double cone $\cO$, i.e. $I''=\cO$, then $\cA(I)=\fA(\cO)$. Similarly  four more intermediate nets and the putative field net $\cF$ are defined. The net $\cF$ realizes the cited simple current extensions.
In Subsection \ref{ss:Chiral versus time zero} the relation between the chiral and the time zero formulation is discussed. The usual d'Alambert formula gives an isomorphic correspondence between the symplectic spaces and the charges in the two cases. In  Subsection \ref{ss:dhr gauge group} we present a detailed description of the twisted and untwisted  automorphisms describing the sectors. Finally, in Subsection \ref{ss:gaugess} the global compact Abelian gauge group $\cG$ is derived as the Bohr compactification of a subspace, isomorphic to $\bR^2$, of the symplectic space of the fields. 
Further details on the braided tensor category of the DHR sectors of $\fA$ will be presented in \cite{fabio4}.   
\section{Weyl Algebras}\label{s:Weyl}
We recall in this section some essential results of the theory of Weyl algebras and fix the general notation, referring mainly to \cite[Section 5.2]{BRII} and \cite{mstv,sla72}.

\smallskip
For $V$ a separable real (topological) vector space, we denote by $\s_V$ a (continuous) \emph{symplectic form} on it, i.e. a $\bR-$bilinear, antisymmetric, real valued, (continuous) form on $V \times V$ and by $(V, \s_V)$ the associated \emph{real symplectic space}. 
The abstract *-algebra on $\bC$ generated by the elements $W(v),\,v \in V $, with product and involution defined respectively by
\beq\label{e:def Weyl algebra}
\barr{ll}
     &W(v)\ W(v') = \ e^{-\frac{i}{2}\s_V (v, v')}\  W(v+v')\,, \\
    & W(v)^* = \ W(-v)
\earr
\eeq
for $v,v' \in V$ is called the \emph{Weyl algebra of $(V,\s_V)$} and indicated by $\cW (V,\s_V)$, or with $\cW_V$ if no confusion arises.

\smallskip
We have to note that, as far as we consider abstract non represented Weyl algebras, it turns out to be needless to specify the topology on  $V$, i.e. we can use the discrete one. Passing to the representations on a Hilbert space, the topology on the (support of the) symplectic space will play its role: Fock representations and, more generally, \emph{regular} representations are typical examples, as we shell better see in the sequel.

\smallskip
The relations (\ref{e:def Weyl algebra}) imply that $W(0)= I$, $W(v)^{-1} = W(-v)= W(v)^*$, i.e. the generators of the Weyl algebra are \emph{formal unitaries} and
\beq\label{e:Weyl algebra}
W(v)W(v') = e^{- i \s_V(v, v')}\  W(v') W(v),\qquad  v,v' \in V\,.
\eeq
The algebra $\cW(V,\s_V)$ is a unital, generally non commutative *-algebra, that is simple iff the symplectic form $\s_V $ is \emph{non degenerate}, i.e. if $\s_V (v, v') = 0$ for all $v \in V$ implies $v' =0$.

\smallskip
In \cite{mstv,sla72} a well established standard theory associates a unique C*-norm to any Weyl algebra, called the \emph{minimal regular norm}. The symplectic form $\s_V$ is non degenerate on $V$ iff a unique C*-norm on $\cW(V, \s_V)$ exists, hence coinciding with the minimal regular one.
We denote by $\csv$ the C*-algebra generated by the Weyl algebra $\cW(V, \s_V)$ in the minimal regular norm, and call it \emph{the (unique) C*-algebra} associated with $\cW(V, \s_V)$.
We term
\beq\label{e:V_0}
N_V:=\,\{v\in V :\forall\, v'\in V,\,\s_V(v,v')=0\}
\eeq 
the \emph{degeneracy subspace} of $V$, so that $\cW(N_V,\s_V)\cong\cW(N_V,0)\subseteq\cW(V,\s_V)$ is the Abelian *-subalgebra generated by $N_V$.  Its completion in the minimal regular norm on $\cW(V,\s_V)$, denoted by $C^*(N_V)$, constitutes the center of $\csv$, i.e. 
\beq\label{e: Weyl center} 
\fZ_V \,:=\, \csv\cap\csv'\,=\,C^*(N_V)\,.
\eeq
$\csv$ is simple iff $N_V=\{0\}$ and in the degenerate case, i.e. $N_V \neq\{0\}$, the minimal regular norm on $\cW_V$ is not the only C*-norm on $\cW_V$.

Clearly, if $V_0:=V/N_V$ and $\s_{V_0}:=\s_V\rest V_0$ (we use the notation $\s_H:=\s_V\rest H$ for the restriction of the symplectic form to a subspace $H\subset V$), the pair  $(V_0, \s_{V_0} )$ is a non degenerate symplectic space and the C*-algebra $\csvn$ it generates is simple. 
The degenerate case is treated in \cite{mstv}, also when $V$ is replaced by an Abelian topological group.

\smallskip
If $\s_V$ is non degenerate and $V$ has a \emph{complexification}, i.e. an operator $J$ such that $\s_V(\cdot, J\cdot)$ is a positive definite form and 
\beq\label{e:J}
 \s_V (Jv, v') = -\s_V ( v, Jv'), \qquad  J^2 =-1, \quad v, v' \in V\,,
\eeq 
we immediately get a pre-Hilbert space structure for $V$, whose inner product is defined from the symplectic form by 
\[
(\cdot,\cdot)_V: =\s_V (\cdot,J\cdot ) +i \s_V(\cdot, \cdot)\,.
\]
Actually such a correspondence between the pair $\s_V,J$ and $(\cdot,\cdot)_V$ is bijective, up to isomorphism, and the complexification is necessary to obtain a pure quasi-free state and a Fock representation for $\cW_V$ (see e.g. \cite{BRII}). This is the usual method, necessary in some sense, to obtain a definite metric Hilbert space representation, for the (observable) algebra of a physical model.
\subsection{Isomorphisms and twisted crossed products of Weyl algebras}\label{ss: isom twist}
We focus in the sequel on two relevant symplectic structures: \emph{isomorphisms} and \emph{twisted compositions} of symplectic spaces; these respectively give rise functorially to isomorphic and twisted crossed products of Weyl and C*-algebras.

\smallskip
A \emph{symplectic morphism} between symplectic spaces is given as a (continuous) map on the spaces, preserving the symplectic forms. An invertible morphism, i.e. an isomorphism, may be defined also in the case of degeneracy as follows
\bdfn\label{d: isomorphism}
Given two symplectic spaces $(V_1,\s_1)$ and $(V_2,\s_2)$, a \emph{symplectic space isomorphism} $\psi: (V_1,\s_1)\to (V_2,\s_2)$ is a continuous isomorphism between $V_1$ and $V_2$ as real topological vector spaces, that preserves the symplectic form, i.e  
\[
		\s_2(\psi(x), \psi(y))=\s_1(x,y), \qquad x,y \in V_1\,.
\]
\edfn
A symplectic isomorphism exponentiates functorially to a Weyl algebras isomorphism between $\cW(V_1, \s_{V_1})$ and $\cW(V_2,\s_{V_2})$ and to a center-preserving C*-algebras isomorphism denoted by
\beq\label{e:Psi}
\Psi: C^*(V_1, \s_1)\longrightarrow C^*(V_2, \s_2)\,.
\eeq
Consider now
\bitem
\item[]
$\catW:=(W(V,\s_V),\f)$, the category of all the Weyl algebras as objects and all the (purely algebraic) isomorphisms  between them as morphisms. 
\eitem
The above discussion may be formalized saying that there exists a \emph{Weyl exponentiation functor} $\bbW$ that realizes an isomorphism of categories between 
\bitem
\item
$\catV:=((V,\s_V), \psi))$, the category of symplectic spaces as objects and symplectic isomorphisms as morphisms;  
\item 
$\bbW(\catV)$, the subcategory of $\catW$, where the morphisms are only $\bbW(\psi)$ for $\psi$ a symplectic isomorphism as in Definition \ref{d: isomorphism}. 
\eitem
Using  the C*-closure of Weyl algebras in the uniquely defined minimal regular norm, both of these categories are isomorph to the following one in C*-context, from which $\bbW(\catV)$ is obtained by a forgetful topology functor:
\bitem
\item
$\bbC^*(\bbW(\catV))$, the subcategory of  Weyl C*-algebras as objects and the unit preserving isomorphisms $\bbW(\psi)$ of $\bbW(\catV)$ as morphisms, extended to the C*-closure. 
\eitem
Notice that the objects in the previously listed isomorphic categories have different algebraic and topological structures, although defined in a natural way starting from $\catV$. A similar natural, physically motivated definition of the representations for the Weyl algebra models is also pursued in the sequel: a typical example is the Fock representation, at least for a Weyl subalgebra and its extension to the whole Weyl algebra.
\smallskip

We may introduce a fourth category, isomorphic to the three above, that is more handy from the point of view of the crossed products theory. 
The objects of this category are called \emph{Weyl algebra groups} and defined as in the sequel: to any (also degenerate) symplectic space furnished with the discrete topology, a Weyl group $\cU(V, \s_V)$ is associated such that 
\beq\label{e:exact seq U(L)}
  1 \to \bT \to \cU(V, \s_V) \to \cU(V) \to 1
\eeq
is a short exact sequence. 
\footnote{
Given a group $G$ an \emph{extension} $E$ of it by another group $N$ is described by the \emph{short exact sequence}
\beq\label{e:exact seq}
  1 \rightarrow N \rightarrow E \rightarrow G \rightarrow 1
\eeq
where $E$ is the set of pairs $(n, s) \in N\times G$ with multiplication law
\[ (n, s)(m, t) := (n\ \b_s (m)\ y(s,t),  s\ t),
\quad (n,s), (m,t)\in N \times G
\]
for $z=(\b, y) : (G, G\times G) \to (\aut N, N)$ the non Abelian  \emph{\em
2-cocycle of the extension} satisfying the equations 
\beq\label{e:2 cocycle}
\barr{c}
  y(s, t) \in (\b_{st},\, \b_s \circ \b_t), \quad s, t \in G \\
 \b_r (y(s, t) ) y(r, st)= y(r, s) y(rs, t), \quad r, s,t \in G\,.
\earr
\eeq
The first equation means that $y$ intertwines the action of $\b_{st}$ and of $\b_s \circ \b_t$, i.e. $y(s, t)\, \b_{st}(n) =\b_s(\b_t(n)) \,y(s,t)$, for every $s,\,t \in G$ and $n\in N$; the second relation is a 2-cocycle multiplicative non Abelian equation.
The extensions $E$ are classified, up to isomorphism, by the 2-cohomology of $G$, with values in 2-category $(Z(N),\aut (N), N)$, where elements in $Z(N)$, the center of $N$, implement identity of $\aut (N)$ of above described cocycles, see \cite{rob77,rob98}.
}
This means that the discrete twisted crossed product $\cU (V,\s_V):=\bT\rtimes_{(\i, y)}\cU(V)$ is an extension of the Abelian formal symbols group $\cU(V)$ on the symplectic space $V$, by the torus group $\bT$ and the 2-cocycle (see \cite{rob98})
\beq\label{e:2-cocy}
z=(\b, y):(\cU(V), \cU(V)\times\cU(V))\to (\aut \bT, \bT)
\eeq
where the action is trivial, $\b\equiv \i$, and the function $y(v,v'):=e^{-\frac{i}{2}\s_V(v,v')}$ is defined by the symplectic form.
Hence, the above announced fourth category is defined by
\bitem
\item
$\bbU(\catV):=(\cU(V, \s_V), \Psi)$, the category of the \emph{Weyl algebra  groups} as objects and the symplectic derived group isomorphisms as morphisms, i.e. $\Psi=\bbW(\psi)$ for $\psi$ as in Definition \ref{d: isomorphism}.
\eitem
A Weyl algebra is hence simply recovered as a discrete crossed product $ \cW(V, \s_V) = \bC \rtimes_{(\i, y)}\cU(V)$. Observe that this is not a semidirect product, eventually defined by a non trivial action $\b$, but a crossed product twisted by the non trivial function $y$.

A useful decomposition in the case of degeneracy is also possible, where  the equation (\ref{e:exact seq U(L)}) is better replaced by an extension making the degeneracy explicit:
\beq\label{e:exact seq deg U(L)}
  1 \to \bT\times \cU(N_V) \to \cU(V, \s_V) \to \cU(V/N_V) \to 1\,.
\eeq 
Here $\bT\times \cU(N_V)\cong Z(\cU(V, \s_V))$ is the center of the group $\cU(V, \s_V)$, and we have $\cU(V/N_V)\cong \cU(V)/\cU(N_V)$.
This extension may be read as the discrete twisted crossed product
\beq
\cU(V, \s_V)=(\bT\times \cU(N_V))\rtimes_{(\i,y)}\cU(V/N_V)\,,
\eeq
where $y$ take value on the $\bT$-part of the normal Abelian subgroup $\bT\times \cU(N_V)$, and the Weyl algebra is also written as $ \cW(V, \s_V) = \cW(N_V) \rtimes_{(\i, y)}\cU(V/N_V)$.

\smallskip
Another simple example of extension is obtained from a \emph{direct sum of symplectic spaces} $(H, \s_H)$ and $(L,\s_L)$ defined by:
\beq\label{e:sympl sum}
(V,\s_V):=(H\oplus L, \s_H + \s_L)\,.
\eeq 
Here we mean that the symplectic form $\s_V$ decomposes according as $\s_V=\s_H \oplus \s_L$, i.e.  $\s_H =\s_V\rest H$ and  $\s_L=\s_V\rest L$, such that $(V,\s_V)\cong (H,\s_H)\oplus(L,\s_L)$ is an obvious symplectic isomorphism that at Weyl algebras level gives 
\beq\label{e:weyl cong}
\cW(V,\s_V)\cong\cW(H, \s_H)\otimes\cW(L, \s_L)\,. 
\eeq
The definition of the C*-maximal tensor product of two C*-algebras gives  the C*-algebra isomorphism 
\beq\label{e:iso split}
\Psi:\csv\longrightarrow \csh\tmax \csl\,.
\eeq
This is easy to obtain because denoting by  
$\norm{.}_H$, $\norm{.}_L$ and $\norm{.}_V$ the minimal regular norms on $\cW(H, \s_H)$,  $\cW(L, \s_L)$ and $\cW(H\oplus L, \s_H \oplus \s_L)$ respectively, for given $a\in \cW(H)$ and $b\in \cW(L)$, on a generic elementary tensor $a\otimes b \in \cW(H)\otimes\cW(L)$ it holds
\[
  \norm{a\otimes b}_{\textrm{max}}=\norm{ab}_V\geq \norm{\Psi(a)}_{\textrm{max}} \norm{\Psi(b)}_{\textrm{max}}= \norm{a}_H \norm{b}_L
\]
being $\norm{\Psi(a)}_{\textrm{max}}= \norm{a\otimes I}_{\textrm{max}}=\norm{a}_H$ and similarly for $b\in \cW(L)$.

We call such a kind of isomorphism for symplectic spaces, or Weyl and associated C*-algebras, a \emph{splitting isomorphism} and a direct sum as in equation (\ref{e:sympl sum}) may also be called \emph{a splitting decomposition of the symplectic space $V$}.
\brem\label{r:tensor}
Observe that if both $(H, \s_H)$ and $(L, \s_L)$ are non degenerate symplectic spaces, the minimal regular norms $\norm{.}_{H}$ and $\norm{.}_{L}$ are unique and the C*-subcross norms on the algebraic tensor product $\csh\otimes\csl$ all coincide, so that 
in this case, it holds (see e.g. \cite{sla72, mstv})
\beq\label{e:tensors}
\csv=\csh\tmax \csl=\csh\tmin \csl\,.
\eeq
\erem
The splitting isomorphisms are trivial examples of the following general construction: let $(H,\s_H)$ and $(L,\s_L)$ be two symplectic spaces, with symbol Abelian groups $\cU(H)$ and $\cU(L)$ and let $\cU(H,\s_H)$, $\cU(L,\s_L)$ be their Weyl algebra groups, defined as in the above equation (\ref{e:exact seq U(L)}). 
Consider the  2-cocycle $(\b, y): (\cU(L), \, \cU(L)\times \cU(L))\to (\aut \cU(H,\s_H),\, \cU(H,\, \s_H))$, defined for the elements  $s=W(l),\,s' = W(l') \in \cU(L)$ and $t=(\z,\,W(h)) \in \cU(H,\, \s_H)$ by
\beq\label{e:azione prodotto}
\b_s (t) = \b_s ((\z,\,W(h)))=(\z \, e^{-i\,\a(h,l)},\,W(h))
\eeq  
where the action $\b$ is given by a (continuous) real valued, $\bR$-bilinear form $\a$, defined on $H\times L$, such that $\a(h,0) = \a (0, l)=0$, and the function $y$ defined by
\beq\label{e:cociclo prodotto}
 y(s,\, s') = ( e^{-i\,\s_L(l,l')/2},\,I_H)\,.
\eeq
To the pair of groups $\cU(H,\s_H)$ and $\cU(L)$ is associated the extension 
\[
e\longrightarrow \cU(H,\s_H)\longrightarrow \cU(H\oplus L, \s_V)\longrightarrow \cU(L)\longrightarrow e 
\]
where $\s_V$ is a symplectic form on $V:=H\oplus L$, defined by
\beq\label{s:forma simplettica}
\s_V ((h,l),\,(h',l')) = \s_H (h,h') + \s_L (l,l') + \a(h,l') -\a(h',l)
\eeq
so that 
\beq\label{e:interact}
\s_{H,L}((h,l),\,(h',l')) :=\a(h,l') -\a(h',l)
\eeq
represents the \emph{interacting content of the non splitting sum}.
 
An extension group is defined from the 2-cocycle $z:=(\b,\,y)$ as above, i.e. in other notation
\beq\label{e:estensione}
\cU(H\oplus L, \s_V) = \cU(H,\s_H) \rtimes_{(\b,y)}\cU(L)= \cU(H,\s_H) \rtimes_z \cU(L)\,.
\eeq 
Explicitly, for generic elements $t=(\z,W(h)),\,t'=(\z',\, W(h'))\in \cU(H, \s_H)$ and $s=W(l), s'=W(l') \in \cU(L)$, the extension group is defined by the product
\beqan
&&((\z,  W(h)),W(l)) \, ((\z', W(h')),W(l'))\,=\\
&&\qquad=\,((\z,\,W(h))\,  \b_s (\z',\,W(h')) \,, W(l) W(l'))\\
&&\qquad=\,((\z \z' e^{-i \a(h',l)}  e^{-i \s_L(l,l')/2} e^{-i\,\s_H(h,h')/2},W(h+h')),  W(l+l'))\,,
\eeqan
by the identity $e=((1, I_H), I_L)$ and the passage to the inverse given by
\[ 
((\z,W(h)), W(l))^{-1}= ((\z^{-1} e^{-i\,\a(h,l)},\,W(-h)),\,W(-l))\,.
\]
In this generality, we can introduce the following
\bdfn \label{d:semidirect product} 
The algebra on $\bC$ associated as group algebra to the extension group $\cU(H \oplus L, \s_V) = \cU(H, \s_H) \rtimes_{(\b,y)}\cU(L)$, where the symplectic form $\s_V$ and the 2-cocycle $(\b,y)$ are defined as above, is called the \emph{twisted crossed product algebra} of the Weyl algebras $\cW(H,\s_H)$ and $\cW(L,\s_L)$. This algebra may also be defined as the Weyl algebra on the symplectic space $(V:=H\oplus L,\s_V)$.
\footnote{
Such a construction from two symplectic spaces is also called the \emph{semidirect product of Weyl algebras} in the literature, see e.g. \cite{her98}.
}
\edfn
In particular cases, such a twisted crossed product of Weyl algebras may be derived from a non splitting decomposition of a symplectic space, as better said in the sequel.

If  $(V,\s_V)$ is a (degenerate) symplectic space and $H$ is a real subspace of it, we denote by 
\beq
H':=\{v\in V: \s_V(v,h)=0,\, h\in H\}
\eeq
the \emph{symplectic complement} of $H$ in $V$ and by
\beq
H^{\bot_{\s_V}}:=\{S\subset V,\, \textrm{linear space}: \, \s_V(s,h)=0,\,s\in S,\, h\in H\}
\eeq
the partially ordered set of \emph{the symplectic subspaces of $V$ disjoint to $H$}. The set $H^{\bot_{\s_V}}$ has maximal element $H'$ and obviously contains the (eventually non trivial) degeneracy subspace $N_V$. 
The decomposition seen in equation (\ref{e:sympl sum}) holds  iff $\s_V(l,h)$ vanishes  for all $l\in L$ and $h\in H$, i.e. introducing the symbol $\bot_{\s_V}$ called \emph{the symplectic disjunction} in $(V,\s_V)$, iff $L\bot_{\s_V} H$.
\smallskip

To construct examples of Weyl algebras products, suppose given a space decomposition $V=B\oplus C$ such that the symplectic form is not splitting, i.e. $\s_V\neq \s_B + \s_C$, and there exists a decomposition of one of the addend as $B=H\oplus N$, with $N\bot_{\s_V} H$ and $C\bot_{\s_V} H$, i.e. $(N\oplus C)\bot_{\s_V} H$. In such a situation we have for the \emph{interacting part of the symplectic form}
\[
\s_{B,C}=\s_{H\oplus N,C}=\s_{N,C}
\] 
and the non splitting contents of such a decomposition of $V$ is confined in the subspace  $L:=C\oplus N\cong N\oplus C$. The Weyl algebra associated to $(V, \s_V)$ is isomorphic to a twisted cross product, in accordance with the following
\bprop\label{p:Weyl crossed product}
Given a symplectic space $(V,\s_V)$ with decomposition 
\[
V=H\oplus N\oplus C,\quad \textrm{with}\quad B=H\oplus N \quad \textrm{and}\quad 
L=C\oplus N\bot_{\s_V} H
\]
there exists a 2-cocycle 
\[
z=(\b,y):(\cU(C),\cU(C)\times \cU(C))\longrightarrow(\aut (\cU(N,\s_N),\cU(N,\s_N)))
\] 
as in equation (\ref{e:2 cocycle}), such that for fixed elements  $s=W(c)$ and $ s'=W(c')$ in $\cU(C)$ an action $\b:\cU (C )\to\aut (\cU (N, \s_{N}))$, is defined by 
\beq\label{e:action L0}
  \b_s (m)= (e^{-i \s_L( c , n)}\z,\, W(n)) = \ad s \,(m)\,,
\eeq
for the element $m= (\z,\, W(n))\in \cU (N,\s_{N})$, and where $y: \cU(C)\times \cU(C) \to \cU(N, \s_{N})$ can be written as 
\[
y(s,\,s'):= (e^{-i\,\s_{C}(c,\,c')/2},\,I)\in \cU(N, \s_{N})\,.
\] 
Such a 2-cocycle gives a twisted crossed product decomposition of the Weyl algebra as   
\beqn
\cW ( V, \s_V ) \,=\, \cW (H\oplus N, \s_H \oplus \s_N) \rtimes_{(\b,\, y)}\cU(C )\,=\,
				\cW(H,\s_H)\otimes \cW(N,\s_N)\rtimes_{(\b,\, y)}\cU (C)\,.
\eeqn
\eprop
\prf
The subspaces $H$ and $L$ in Definition \ref{d:semidirect product} have to be respectively identified with $H\oplus N$ and $C$ in the case at hand. According to this, in the equation (\ref{s:forma simplettica}) we have to read $\a(h,c)=\a(n,c)=\s_L(n,c)$ for $h\oplus n\in H\oplus N, \,c\in C$, and the symplectic form $\s_V$ decomposes as $\s_V=\s_H \oplus\s_L$, where $L=C\oplus N$ and $\s_L=\s_V\rest L$.
\qed
Observe that the subspace $N$ may be thought, is some sense, as being \emph{in common} between the  symplectic subspaces $B$ and $L$, and that the Weyl elements defined from the subspace $C$ have a non trivial action on Weyl elements defined from $N\subset B$, by the evaluation of $\s_L$.

\smallskip
We end this section with some broad ideas about the formalization of physical model by Weyl algebras. In all generality, a \emph{simple current extension} of Weyl algebras is essentially described by a crossed product of Weyl algebras, along the following scheme. 
The charge carrying fields are defined starting from a symplectic space
\beq\label{e:model gen}
(V_f,\s_f)\, =\, (V_a \oplus N\oplus C,\s_f)\,.
\eeq
Here, we have to read the subspace decomposition $V=B\oplus C=(V_a\oplus N)\oplus C$  as in Proposition \ref{p:Weyl crossed product} above, with $H=V_a$. The symplectic form $\s_f$ may not split in the sum $\s_B \oplus \s_C$, for $\s_B=\s_f\rest B$ and $\s_C=\s_f\rest C$, but if  $N\in V_a^{\bot_{\s_f}}$ the field Weyl algebra isomorphically can be written as
\beq\label{e:ph crossed}
\cW(V_f,\s_f)\,=\,\cW (V_a\oplus N,\s_a \oplus \s_N)\rtimes_{(\b,\, y)}\cU(C)\,.
\eeq
In these decompositions, $V_a$ has the meaning of  the symplectic space for the observables algebra, for which a (regular positive metric) Fock space representation $\p_a$ exists. The representation $\p_f$ for the field algebra $\cW_f$ is in general a non-regular extension of $\p_a$, as it happens in a non rational model, and $\cU(C)$ plays the role of \emph{the charge group} of the theory. Observe that this description is more general that the one treated in Proposition \ref{p:Weyl crossed product}, where also $C\in V_a^{\bot_{\s_f}}$ was assumed.

Hence the Weyl algebra models may be classified on the basis of the different specific properties in the above space decomposition (\ref{e:model gen}) and the algebraic ones in equation (\ref{e:ph crossed}), such as the dimension of $C$ and $N$ as real linear spaces, the evaluation of $\s_V$ when restricted to $C$ and $N$, and so on. We will see two different examples below. 

As a final remark, observe that the purely algebraic constructions above are shown to entails some general functorial features passing to representations, that are also relevant for the nets of von Neumann algebras, defined from localized symplectic subspaces of a given symplectic space.
\subsection{Representations}\label{ss:reps}
We summarize some general results about the representation theory of a Weyl algebra $\cW(V,\s_V)$ and its associated C*-algebra $\csv$, see \cite{sla72,mstv} for details:
\bitem
 \item
every positive linear functional on $\cW(V, \s_V)$ is continuous with respect to the minimal regular norm and extends to a unique positive, continuous linear functional on $\csv$;
 \item
 every representation $\p$ of the Weyl algebra $\cW(V, \s_V)$ on a Hilbert space $\cH_\p$ extends to a representation of $\csv$, on the same Hilbert space;
 \item
 every *-automorphism on $\cW(V, \s_V)$ extends uniquely to a *-automor\-phism of $\csv$.
\eitem
In the sequel we show the relation between the twisted crossed product characterization of Weyl algebras, introduced in the last section, and some factorization properties of their representations. 
We begin from the simplest situation, the split case of equation (\ref{e:weyl cong}) or (\ref{e:iso split}), by the following
\blemma\label{p:splitting}
Let $(V=H\oplus L, \s_V=\s_H \oplus \s_L)$ be a direct sum of symplectic spaces with Weyl algebra $\cW(V,\s_V)\cong \cW(H,\s_H)\otimes \cW(L,\s_L)$ as above. Then
\bdes
\item[i)]
if $(\p_{\o_H} ,\cH_{\o_H},\O_H )$ and $(\p_{\o_L} ,\cH_{\o_L} ,\O_L )$ are the GNS representations associated to the states $\o_H$ and $\o_L$ on $\cW_H$ and $\cW_L$ respectively, then the unique product state $\o$ and its GNS representation $\p_\o$ is canonically defined for the Weyl algebra $\cW_V=\cW(H\oplus L, \s_H \oplus \s_L)$ by the (spatial) tensor product as 
\[
(\p_\o ,\cH_\o ,\O ) = (\p_{\o_H} ,\cH_{\o_H},\O_H )\otimes (\p_{\o_L} ,\cH_{\o_L} ,\O_L )\,;
\]
\item[ii)] 
if $(H,\s_H)$ and $(L,\s_L)$ are non degenerate symplectic spaces or if (for example) $\s_V\rest L=\s_L$ vanish, i.e. if $\cW(L,\s_L)=\cW(L)$ is Abelian, then   
\[
\big(\p_\o(\csh \tmax \csl))''=\p_{\o_H}(\csh\big)''\otimes\p_{\o_L}(\csl)''\,,
\]
where the latter means the tensor product of the von Neumann algebras $\p_{\o_H}(\csh)''$ and $\p_{\o_L}(\csl)''$. 
\edes
\elemma
\prf
i) $\p_\o$ is obtained as the GNS representation of the \emph{product state} $\o:=\o_H\otimes\o_L$ on the C*-algebra $C^*(H,\s_H)\tmax C^*(L,\s_L)$, i.e. from the state defined by
\[
\o(A\tmax B)\,=\,\o_H(A)\,\o_L(B)\,,\qquad A\in \csh\,,\quad B\in \csl\,.
\]
Here $\tmax$ assures for the product state $\o$ a well behaved passage to the representation $\p_\o$ of $\csh\otimes \csl$ on the Hilbert space $\cH_\o=\cH_{\o_H}\otimes\cH_{\o_L}$, obtained as the spatial tensor product of the GNS representations $\p_{\o_H}$ of $\csh$ and $\p_{\o_L}$ of $\csl$ (see e.g. \cite[Theorem IV.4.9]{TI} or \cite[Proposition 11.1.1]{KRII} for details).\\
ii) 
The results follow directly from \cite[Theorem IV.4.13]{TI} and the identity 
\[
\csh \tmax \csl=\csh \tmin \csl\,.
\] 
This equality, in the case of non degenerate subspaces is given by the equation (\ref{e:tensors}). In the second case, if 
$\csl$ is Abelian, hence nuclear, it is a well known consequence.   
\qed
An elementary example of item ii) in above Lemma \ref{p:splitting}, is given by a splitting isomorphism of symplectic spaces with $L=N_V$, the degenerate subspace of $V$, and $H=V/N_V$. 

\smallskip
Passing to the non splitting situation, the factorization of representations we are interested in, is described by the following general result, also related to one in \cite{her98} 
\bprop\label{P:factor state weyl case}
Let  $(V=H\oplus L, \s_V= \s_H +\s_L +\s_{H,L})$ be a symplectic space decomposition as in Definition \ref{d:semidirect product} and equation (\ref{e:estensione}), such that the Weyl algebra $\cW(V,\s_v)$ is not splitting, i.e.  the real form $\a$ that defines through the equation (\ref{e:interact}) the interacting part $\s_{H,L}$ of the symplectic form $\s_V$ is non trivial. Then, for given $\o_H$ and $\o_L$ two states on $\cW(H,\s_H)$ and $\cW(L,\s_L)$ respectively, the linear functional on $\cW(V,\s_V)$ defined for $v=h\oplus l\in H\oplus L=V$ by 
\beq\label{e:omega 1}
 \o (W(v)) := \o_H (W(h)) \, \o_L (W (l))\,, 
\eeq
is positive, i.e. is a state on $\cW(V,\s_V)$, if  $\o_L(W(l))=0$ for $l\notin H^\bot\cap L$. 
In particular, if  $H^\bot\cap L=\{0\}$, i.e. if $\a$ is non trivial on any subspace of $L$, such a condition is also necessary, i.e. if  $H^\bot\cap L=\{0\}$, $\o$ is a state on  $\cW(V,\s_V)$ iff
\beq\label{e:condition 2 weyl}
\o_L (W(l)) =0 \qquad \textrm{for}  \qquad  l \neq 0\,.
\eeq  
If $H^\bot\cap L=\{0\}$ and the condition (\ref{e:condition 2 weyl}) holds, the state $\o$ is faithful iff so is the state $\o_H$. Respectively changing $L$ with $H$. 
\eprop
\prf
To verify the hermiticity of $\o$, we may restrict to an element $A=W(l)W(h)$ for $l\in L$ and $h\in H$, so that such a property for $\o$ holds iff  
\beqan
	\o (A) &=&\o_{H} (W(h))\, \o_L (W(l))\,  =\, \ov{\o (A^* )} = 
	\ov{\o (W(l)^* \b_{W(l)}(W(h)^* ))}\\
	 &=& \o_H (\b_{W(l)} (W(h)))\, \o_L(W(l))\, 
\eeqan
i.e., by the definition of the action of $\b$ in equation (\ref{e:action L0}), iff
\[ 
\o_H(W(h))\,\o_L (W(l))  \, \big( 1 -e^{i\, \a(h,l)}\big) =0\,.
\]
Hence the hermiticity holds if $\o_L(W(l))=0$ for $l\notin H^\bot\cap L$. In particular, if $H^\bot\cap L=\{0\}$, the hermiticity holds iff the condition (\ref{e:condition 2 weyl}) is satisfied.

To show the positivity, observe that any element $A\in\cW_V$ is written  for $l_i\in L$, $h_i\in H$ and $a_i\in \bC$ with $l_i\neq l_j$ for $i\neq j$, as a finite sum
\beq\label{e:summa}
A=\sum_{1\leq i\leq n} a_iW(h_i)W(l_i)\,.
\eeq
Hence we obtain 
\beqa\label{e:aa*}
AA^*&=&\sum_{1\leq i\leq n} |a_i|^2 \,+\sum_{1\leq i<j\leq n} a_i \ov{a}_j W(h_i)W(l_i)W(l_j)^*W(h_j)^*\,+\, adj \nonumber\\
    &=&\sum_{1\leq i\leq n} |a_i|^2 \,+\sum_{1\leq i<j\leq n} a_i \ov{a}_j e^{i\a_{ij}} W(h_i-h_j)W(l_i-l_j)\,+\, adj 
\eeqa
where $\a_{ij}=\frac{1}{2}[\s_H(h_i,h_j)+\s_L(l_i,l_j)+ \s_V(h_j,l_i-l_j)]$. 
Applying the functional $\o$ gives
\beqa\label{e:notriv}
\o(AA^*) &=& \sum_{1\leq i\leq n} |a_i|^2 \,+\sum_{1\leq i<j\leq n} 2 \Re\{ a_i \ov{a}_j e^{i\a_{ij}} \o_H(W(h_i-h_j)) \o_L(W(l_i-l_j))\}\nonumber \\
         &=& \sum_{1\leq i\leq n} |a_i|^2 \,+\sum_{\genfrac{}{}{0cm}{3}{1\leq i<j\leq n}{l_i-l_j \in H^\bot \cap L}} 2 \Re\{ a_i \ov{a}_j e^{i\b_{ij}}\}\nonumber\\
         &\geq& \sum_{\genfrac{}{}{0cm}{3}{1\leq i<j\leq n}{l_i-l_j \in H^\bot \cap L}} |a_i|^2 +\sum_{\genfrac{}{}{0cm}{3}{1\leq i<j\leq n}{l_i-l_j \in H^\bot \cap L}} 
         2 \Re\{ a_i \ov{a}_j e^{i\b_{ij}}\} \,= \sum_{\genfrac{}{}{0cm}{3}{1\leq i<j\leq n}{l_i-l_j \in H^\bot \cap L}} |b_i|^2\,\geq\,0  
\eeqa 
where we used  the condition $\o_L(l)=0$ for $l\notin H^\bot \cap L$, redefined $\b_{ij}$ from the phases that arise from the evaluation of the states, and the last equality holds because we may call $b_i=a_i e^{i\b_i}$ for some $\b_i$ such that $\b_{ij}=\b_i-\b_j$. If we suppose that the property (\ref{e:condition 2 weyl}) holds true for $L$, we only have $\o(AA^*)=\sum_{1\leq i\leq n} |a_i|^2$, so the faithfulness trivially follows in this case.
\qed
Observe that the above defined state $\o$ is a product state on the tensor product algebras $\cW_H\otimes \cW_L$ in the sense of Lemma \ref{p:splitting}, only if the linear form $\a$ trivializes, i.e. if $H^\bot \cap L =L$. 

In particular, the Proposition \ref{P:factor state weyl case} applies to the symplectic space decomposition as in Proposition \ref{p:Weyl crossed product}. For this case, given the two states $\o_H$ on $\cW(H, \s_H)$ and $\o_L$ on $\cW(L, \s_L)$, we define in a canonical way as in Lemma \ref {p:splitting},  the product state  
\beq\label{e:omega p}
\o_p := \o_H \otimes \o_L
\eeq
on the tensor product Weyl algebra $\cW(H, \s_H)\otimes \cW(L, \s_L)$ and denote by $\p_p := \p_H \otimes \p_L$ its GNS representation on the Hilbert space $\cH_p = \cH_H \otimes \cH_L$.  

\smallskip
Now let $\o_{H\oplus N}$ be an extension of the state $\o_H$ from the Weyl algebra  $\cW(H,\s_H)$ to the Weyl algebra $\cW(H\oplus N, \s_{H\oplus N})$, for an extension $\s_{H\oplus N}$ of the symplectic form $\s_H$ of $H$ to $H\oplus N$. 
For $L=N\oplus C$, let denote by $\o_C$ the restriction of a state $\o_L$ of $\cW(L,\s_L)$ to the subalgebra $\cW(C, \s_L\rest C)$. 
A result relating the states on a twisted crossed product decomposition (Proposition \ref{P:factor state weyl case}) and a splitting Weyl algebra (Lemma \ref {p:splitting}), is obtained confining the interacting part $\s_{H,L}$ of the symplectic form $\s_V$ to the subspace $N$, that is \emph{in common} between $H$ and $L$.

Hence we give the following result, where indeed, referring to Proposition \ref{P:factor state weyl case}, $H$ is replaced by $H\oplus N$ and $L$ by $C$:
\bprop\label{P:state coincidence}
Consider $V=H\oplus L = H\oplus N\oplus C$ as in notation above, so that the state  $\o$ in equation (\ref{e:omega 1}) is well defined. Then the following are equivalent:
\bdes
\item[i)]
$\left\{ \barr{ll}
\o_{H\oplus N} (W(x)) = \o_H (W(h))\,, \quad \quad \quad &x= h\oplus n \in H\oplus N\,,\\
 \o_L (W(l)) = \o_C (W(c))\,, &l= n\oplus c \in N\oplus C=L\,;
 \earr
 \right.$
 \item[ii)]
the states $\o_p$ and  $\o$, the first defined as in (\ref{e:omega p}) with extension and restriction as above, and $\o$ defined as in Proposition \ref{P:factor state weyl case}, coincide on $\cW(V, \s_V)$ and on its C*-algebra $\csv$.
\edes    
\eprop
\prf
i) $\Rightarrow$ ii) is obtained by a simple calculation from the definitions, for $v = h\oplus l=h\oplus n\oplus c \in H\oplus N\oplus C= V$ and $x =  h\oplus n \in H\oplus N$
\[ \o_p (W(v)) := \o_H (W(h))\otimes \o_L (W(l)) =  \o_{H\oplus N} (W(x))\, \o_C (W(c)) =: \o (W(v))\,. 
\]
ii) $\Rightarrow$ i) It suffices to use $c =0$ in the preceding calculation to obtain back the first relation in i) and analogously, using $h =0$, for the second.
\qed
Notice that for the element $h=0$ in $H$ we obtain from the condition i) in the Proposition above that the state $\o_N:=\o_{H\oplus N}\rest N$ of $\cW_N $ is such that $\o_N(W(n))=1$ for every $n\in N$.
\brem
1) Consider a symplectic space decomposition as in Proposition \ref{p:Weyl crossed product} and refer to the physical model discussion about equation (\ref{e:ph crossed}). If $C$ represents the charge group for $V_f$ and $H\oplus N$ is related to the symplectic space of observables, then the triviality of $\a$, i.e. of $\s_{N,C}$, is equivalent to a trivial action of the sector (auto)morphisms on the observables, that means a trivial superselection structure of the model.
The condition in equation (\ref{e:condition 2 weyl}) instead, defines the state $\o_L$ in a unique way as a non-regular state (see Definition \ref{d:non-regular} below).\\
2) The obtained results actually extend the ones of Herdegen in \cite{her98}, where a crossed product of a Weyl algebra by a CAR algebra is treated: in fact, as observed by Slawny in \cite[Section 3.10]{sla72}, a CAR algebra may always be written as a Weyl algebra with a non degenerate symplectic form.\\
3) The Proposition \ref{P:state coincidence} suggests that, for twisted crossed products as above, it is the non-regular representations of $\cW(V,\s_V)$ that is of interest, with $C$ the \emph{non-regular subspace} and $N$ the \emph{regular subspace} of $V$, see Definition \ref{d:non-regular} below.  This always leads to non rational models, with uncountably many superselection sectors, in the sense of \cite{lx04}.
\erem
Further general results for twisted crossed products and non-regular representations of CCR algebras of a generic locally compact group, along the lines of \cite{gru97} are also possible, see \cite{fabio03}.
\subsection{Elementary Weyl algebra in non-regular representation}
\label{ss:Elementary Weyl algebras}
In the above subsection we pointed out the utility of non-regular representations when a  field algebra model is written as a cross product of Weyl algebras. A regular (Fock space) representation is used instead on the observable part  of the model, also called in literature \emph{the physical part}, and a non-regular one on the charge part, see e.g. \cite{ams93}. We recall hence the following basic   
\bdfn\label{d:non-regular}
A representation $\p$ of a Weyl algebra $\cW=\cW(V,\s_V)$ is said a \emph{regular representation} if for all $v\in V$ the one parameter group
\beq\label{e:regular repp}
\l \longmapsto \p (W(\l v)),\, \quad \l \in \bR,\, v\in V
\eeq
is weakly (and strongly) continuous.
If there exist a non trivial subspace $V_1\subset V$ such that the map in equation (\ref{e:regular repp}) is not weakly continuous, then the  representation $\p$ is said to be \emph{non-regular}. The maximal such subspace is said \emph{the non-regular subspace} of $\p$. A state $\o$ whose GNS representation $\p_\o$ is regular (non-regular) is also said to be regular (non-regular).
\edfn
Observe that in general a Weyl algebra is not norm continuous, being  $\|W(v_1) -W(v_2)\| = 2$ if $v_1 \neq v_2$ for $v_1, v_2 \in V$.
The strong continuity of the regular representations of the Weyl algebras, that is needed for their relevant physical content, is in some sense also necessary to obtain good representation of the associated CCR algebra, see \cite[Section 3.3]{rob98} for details.

\smallskip 
We now discuss in detail the simplest, but relevant, case of non-regular representation for the symplectic space $L\cong \bR_d^2$, in order to use it as a \emph{building block} for the superselection theory of physical models.

Let $C\cong N\cong \bR_d$ be copies of the additive group of reals, furnished with the discrete topology; consider the symplectic space $(L,\s_L)$ where $L=C\oplus N\cong \bR^2_d$ and the symplectic form is the usual non degenerate one on $\bR^2_d$, i.e. for a pair of elements $l=(c,n),\, l'=(c',n')\in L$ we have
\[
W_L(l)\ W_L(l') = \ e^{-\frac{i}{2}\s_L(l,l')}\  W_L(l+l')\,, \qquad \s_L(l,l')= (cn' -c'n)\,.
\]
The Weyl algebra $\cW_L$ associated to $(L,\s_L)$ is hence that of a quantum system with one degree of freedom; its irreducible regular representations, and those of the associated C*-algebra $\csl$, are well described by the Stone-von Neumann uniqueness theorem, see e.g. \cite{BRII}.

To consider a particular non-regular representation of $\cW_L$, following for example \cite[Section 2]{tn92}, let $\o_L$ be the functional on $\cW_L$ defined by
\beq\label{e:state NT}
\o_L(W_L(l))=\o_L (W_L((c,n)))= \d_{c, 0}\,, \qquad  l=(c,n) \in L\,.
\eeq
This is the unique  tracial state on $\cW(L,\s_L)$, up to the change of $C$ and $N$, and it turns out to be not faithful because  
\[
\o_L \big(( I-W_L ((0,n))) ( I -W_L ^* (( 0,n)))\big)=0\,,  \qquad (0,n)\in L\,.
\]
However, if we denote  the GNS triple of $\o_L$ by $(\p_L,\cH_L,\O_L)$,
we observe that $\p_L$ is faithful because $\wL$ is simple, as the degeneracy subspace of $L$ is trivial, i.e. $N_L=\{0\}$.
The GNS Hilbert space $\cH_L$ is non separable and can be derived from the total set of vectors obtained as
\[ 
|c, n\rangle := \p_L (W_L ((c, n)))\,\OL\,, \qquad (c, n) \in L\,.
\]
The action of a represented Weyl element $\p_L (W_L ((c,n)))$ on the generic vector $|c',n'\rangle\in \cH_L$, for $(c,n), \, (c',n')\in L$, reads as
\beq\label{e:weyl op action}
 \p_L (W_L ((c,n)))|c',n'\rangle = 
					e^{-\frac{i}{2} \s_L ((c,n), (c',n'))} |c+c', n+n'\rangle\,.
\eeq
Moreover, if $|c,n\rangle$ and $|c',n'\rangle$ are two elements of $\cH_L$, their scalar product is equal to 
\beq\label{e:inner prod weyl} 
\langle c,n |c' ,n'\rangle = \d_{c,c'}\,  e^{\frac{i}{2} \, c(n'-n)}\,,
\eeq
so that $\langle 0,n |0 ,n'\rangle = 1$. It is hence possible to identify  the vectors $|c,n\rangle$ for every $n \in N$ with the vector $|c,0\rangle$, up to a phase, giving 
$|c, n\rangle = e^{\frac{i}{2} cn } |c,0\rangle$ and $\langle c,n |c',n'\rangle = e^{\frac{i}{2} c(n'-n) } \d_{c,c'}$.
Because of this identification, we simply write $|c\rangle:=|c,0\rangle$. In particular this is possible for $c =0$, so that the following relations hold
\beq\label{e:omegaL}
\OL = |0, 0\rangle= |0, n\rangle =: |0\rangle\,,  \qquad \textrm{for all }n\in N\,.
\eeq
From the above discussion, the Hilbert space $\cH_L$ may also be written as 
\beq\label{e:H_L}
\cH_L = \bigoplus_{c\in C} \bC\, |c\rangle\cong \ell^2 (C) \cong \ell^2 (\bR_d)\,, 
\eeq
$\ell^2 (\bR_d)$ being  the space of square summable maps from $\bR_d$ into $\bC$, i.e.  an element $\x\in\ell^2(\bR_d)$ is a function supported on a countable
subset $D_\x$ of $\bR$ such that $\|\x\|:=\sum _{d\in D_\x}\abs{\x(d)}^{2}<\infty$. 
According to the equation (\ref{e:inner prod weyl}), the inner product on $\ell_2(\bR_d)$ is equivalently given by
\[
\langle \x,\e \rangle = \sum _{d \in D_\x\cap D_\e}\ov{\x(d)}\e(d)\,,\qquad \x,\,\e\in \ell^2 (\bR_d)\,. 
\]
Denoting for every $c\in \bR$ by  $\d_c\in\ell^2 (\bR_d)$ the characteristic function of the subset $\{c \}\subset \bR$, identifying $|c,0\rangle$ with $\d_c$ gives the isomorphism in above equation (\ref{e:H_L}).

\smallskip
The representation $\p_L$ turns out to be irreducible and non-regular, not being  strongly continuous in first coordinate $c\in C$ of elements $W_L(c,r)$ but only in the second.
The von Neumann algebras $\cL:=\p_L (\csl)''$ is irreducibly represented on $\cH_L$ by $\p_L$.
 
Moreover, we may consider the Weyl elements $W_L(c,0),\, c\in C$ and $W_L(0,n),\, n\in N$, and the Abelian C*-subalgebra of $\csl$ they generate, respectively $C^*(C, 0)=:\csco$ and $C^*(N, 0)=:\csn$; finally if $\cC:=\p_L(\csco)''$ and $\cN:=\p_L(\csn)''$ are the associated Abelian von Neumann subalgebras of $\cL$ they define, the properties of these unitaries and algebras are collected in the following
\bprop\label{p: elem weyl rep}
Under the above definitions we have:
\bdes
	\item[i)] 
	if $c\neq 0$ and $n\neq 0$ then the spectrum of $W_L((c,0))$ is $\bT$ and equals both the spectrum of $W_L((0,n))$ and the discrete spectrum of $\p_L(W_L(0,n))$;
	\item[ii)]
	there is a self-adjoint operator $\Phi_N$ on $\cH_L$ such that for any $n\in N$ we have
	$\p_L(W_L(0,n))=\exp \{ \frac{i}{2}\,n\,\Phi_N\}$, i.e. $\Phi_N$ is the generator of the one parameter group associated to $N$;
	there is no self-adjoint operator $\Phi_C$ on $\cH_L$, generating a one parameter group associated to $C$. In particular, for all $c\in C$ we have
\beqn
\Phi_N|c,0\rangle\,=\,-i\lim _{n\to 0}n^{-1}(\p_L(W(0,n))-I_L)|c,0\rangle 
						 \,=\,-i\lim _{n\to 0} n^{-1} (e^{inc} -1)|c,0\rangle=c|c,0\rangle\,;
\eeqn
	\item[iii)] $\csco\cong C(bC)\cong C(b\bR)\cong C(bN)\cong \csn$, where $C(b\bR)$ indicates the C*-algebra of the continuous functions on  the Bohr compactification $b\bR$ of $\bR$, i.e. is the algebra of the almost periodic functions on $\bR$, i.e. the group C*-algebra of the compact group $b\bR$;
		\item[iv)] $\csl=\csn\rtimes_{(\i,y)}\cU(C)$, i.e. $\csl$ is the twisted crossed product of $\csn$ by the discrete additive group $C$, with 2-cocycle $(\i,y)$, defined from the symplectic form $\s_L$;
	\item[v)] every non zero vector in $\cH_L$ is cyclic for the von Neumann subalgebra $\cC$; $\cN$ is a spectrally multiplicity free von Neumann subalgebra of $\cL$. In particular, $\cC$ and $\cN$  are maximal Abelian in $\cB(\cH_L)$;
	\item[vi)] $\cL:=\p_L(\csl)''$ is a type $II_1$ factor.
\edes
\eprop
\prf
i), ii) and iii) are standard results, see for example \cite{sla72} and also \cite{hal04}.\\
For  iv) see the above Proposition \ref{p:Weyl crossed product}.\\
v) Considering the total set of vectors $|c\rangle,\, c\in C$ and the action of $\cC$ on them as read off from equation (\ref{e:weyl op action}), cyclicity for $\cC$ is trivial, and the maximality is a classical result. For the subalgebra $\cN$, consider for every $c\in C$, the spectral measures $\m_c$ defined on  $\widehat{\csn}\cong bN$, the Gelfand spectrum of $\csn$, and the Baire set $X_c = bN\setminus c$. These data satisfy the hypothesis of \cite[Corollary 5.2]{cav98} that characterizes $\p_L$ as a \emph{spectrally  multiplicity-free representation} of $\csl$, a stronger condition for $\cN$ to be maximally Abelian in $\cB(\cH_L)$.\\
vi) The von Neumann algebras $\cL$ is equal to the represented discrete crossed product algebra $\cN\rtimes_\a \G$, where $\G:=\p_L(\cU(C))$ denotes   the discrete group $\cU(C)$ represented on the Hilbert space $\cH_L$. The automorphic action $\a$ defined by $\a_c:=\ad W(c),\,c\in C$ on the maximal Abelian *-algebra $\cN$ is given by the $2-$cocycle $(\i,y)$, i.e. through the symplectic form $\s_L$. It easy to see that $\a$ acts ergodically on $\cN$, so that $\cL$ is a factor. 
Moreover, the state $\o_L$ is a tracial state on $\cL$ so that the algebra is of finite type, and is of type $II_1$ being not finite dimensional.
\qed 
\brem
1) 
The elements $\p(W((c,0)))$ for $c\in C$, may be considered as operators carrying the charge $c$. The elements $\p(W((0,n)))$ for $n\in N$ act as different phases, on the distinct charge subspaces, in the charge decomposition of $\cH_L$ given by equation (\ref{e:H_L}).\\
2) The physical idea is to profit of  a symplectic space $L=C\oplus N$ with $C\cong \bR_d^n \cong N$ for some natural $n$, such that $C$ represents the additive group of the charges  and its associated dual group $\dC\cong b\bR\cong bN$ the internal (gauge) symmetry group of the model. The non-regular representation of $\cW_L$ with $C$ the non-regular subspace of $L$ and obtained as the $n$-th tensor product of the representation $\p_L$ above, may be used to manage the twisted crossed product of Weyl algebras in a defining state satisfying the requirements of Propositions \ref{P:factor state weyl case} and \ref{P:state coincidence}.\\
3) 
From a mathematical point of view, a similar construction  may also be given for the Weyl algebras $\cW(L, \s_L)$ where now $L=C\oplus N$, for $C$ an Abelian discrete group and $N$ the locally compact group such that its Bohr compactification $bN$ is equal to $\dC$, the dual group of $C$, i.e. the space of \emph{all} arbitrary characters of $C$, see \cite{mstv, sla72, gru97, fabio03}.  
\erem
\section{The Streater and Wilde model}\label{s:SW model}
We begin in this section to analyze the scalar massless free Bosonic quantum field, on the 1+1-dimensional Minkowski spacetime, along the lines of the twisted crossed products described in the previous Section \ref{s:Weyl}. General references are \cite{sw70, hl82} and a recent more physical introduction is \cite{dm04}.

\smallskip
With respect to the original formulation on chiral left/right light lines of the Streater and Wilde paper we prefer a time zero formulation, i.e. giving Cauchy data on the time zero real line, because the classification of the superselection sectors, e.g. their different (solitonic) origin, is more clear in this approach, as it will be shown in Subsection \ref{ss:dhr gauge group}.

In contrast to the same model in higher dimension it is well known that in the 1+1-dimensional case infrared infinities occur, so that the observable algebra may be defined in a physical Fock space representation only after furnishing some supplementary constrain conditions reducing the algebra of non observable fields.

\smallskip 
If we write $\cS := \cS_{\bR}(\bR)$ for the Schwartz space of real valued rapidly decreasing functions on the real line, the cited constrains are realized in the Weyl formulation as a restriction of the test function space $\cS$ to a subspace of functions with vanishing Fourier transform at zero momentum, i.e. $\tilde{f}(0):=\int\,f(x) dx=0$, for $f\in \cS$.
\footnote{
Considering localized subspaces of test function space as defining local algebras, we may also replace  $\cS$ by the space of smooth functions with compact support. 
}
This condition is also written for short as $f=\partial g$, for some $g\in \cS$, and is required for smearing the functions of the observable fields at time zero, not for their Lagrangian associated time zero momentum. For this reason, the observable theory is also called in the literature \emph{the  theory of the potential of the field}. 
\subsection{Weyl algebras for the Streater and Wilde model}\label{ss:WSW model}
The symplectic spaces of the model are introduced by the following restriction and extensions of the function space $\cS$:
\beq\label{e:Schwartz spaces}
\barr{l}
\dS:=\{ f \in \cS : f=\partial g ,\  g \in \cS \}\,,\\\\
\duS:=\{ f \in C^\infty (\bR) :\, \partial f \in \cS \}\,,\\\\
\duzS:=\{ f \in \duS :\, \lim_{x\to-\infty} f(x) = \lim_{x\to+\infty} f(x)\}\,,\\\\
\duqS:=\{ f \in \duS :\, \lim_{x\to-\infty} f(x) = - \lim_{x\to+\infty} f(x)\}\,.
\earr
\eeq
Denoting by $\bR_d$ the constant functions on the time zero line, we have the following additive group quotients
\beq\label{e:quotient S}
\cS=\duzS/\bR_d\qquad \textrm{ and }\qquad\duqS=\duS/\bR_d\,,
\eeq 
and the inclusions
\beq\label{e:inclusions S}
\dS\subset\cS,\qquad\cS\subset\duzS\subset\duS\qquad \textrm{and}\qquad \cS\subset\duqS\subset\duS\,.
\eeq
From the linear spaces in equations (\ref{e:Schwartz spaces}) and because of the inclusions in  (\ref{e:inclusions S}) we define properly included (symplectic) spaces as follows:
\beq\label{e:def inclusion symplectic}
\barr{ccccc}
V_a:=\dS\oplus\cS&\subset&V_b:=\dS\oplus\duzS&\subset&V_c:=\cS\oplus\duzS\\\\
 \cap&&\cap & & \cap \\\\
 V_q:=\dS\oplus\duqS &\subset&V_e:=\dS\oplus\duS&\subset&V_f:=\cS\oplus\duS\,.
\earr
\eeq
Moreover, because of the quotient maps in equations (\ref{e:quotient S}) we have, for $N\cong \bR_d$
\beq\label{e:trivial}
V_b=V_a\oplus N\qquad \textrm{and}\qquad V_e=V_q\oplus N\,.
\eeq
The symplectic spaces associated to the above support spaces are defined as follows: for $F=\fzu$ and $G=\gzu$ elements in $V_a$, a (non degenerate) symplectic form $\s_a$ is defined by 
\beq\label{e:sympl form f}
\s_a (F, G) = \int_\bR \, (f_0 g_1 - f_1 g_0 )\, dx\,.
\eeq
The following standard procedure gives a pre-Hilbert space from the symplectic space $V_a$,
defining it on the mass shell (i.e. the positive light cone in momentum space), and leads to a Fock space representation of the associated observable algebras. 
Let $T_a$ be the map defined by
\beq\label{e:T_a}
\barr{rcll}
    T_a :& V_a &\longrightarrow &\fH_a := L^2(\bR,\,dp)  \\
    & F = \fzu &\longmapsto &\o^{-1/2} \tilde{f}_0 +i\, \o^{1/2}\tilde{f}_1\,, \qquad \textrm{for}\quad\o := |p|\,,
\earr
\eeq
from the real space $V_a$ to the real subspace  $T_a (V_a)$ of $\fH_a$, whose closure in $\fH_a$ constitutes the real subspace $H_a:= \ov{T_a (V_a)}$ of $\fH_a$, i.e. the \emph{one particle Hilbert space} of the observables of the model, see \cite{sw70,hl82} for details. 
\footnote{
Because the map $T_a$ is an isomorphism of symplectic spaces, the pair $(T_a(V_a), \s_a)$ itself is often regarded as the symplectic space of the observables of the model.
}
The scalar product on $H_a$ is explicitly given, for two elements obtained from $F, G \in V_a$, by
\beq\label{e:H_a scalar product}
(T_a F, T_a G)_{H_a} := \int \,( \o^{-1} \bar{\tilde{f}}_0 \, \tilde{g}_0   + \o \bar{\tilde{f}}_1 \, \tilde{g}_1 )\, dp\,+\, 
					i \int \,(  f_0 \, g_1  - f_1 \, g_0 )\, dx\,.
\eeq
For $F=\fzu\in V_a$, a quasi-free state on the algebra $\cW(V_a,\s_a)$ is given by
\beqa\label{e:Fock state}
\o_a (W_a(F)) &:=& \exp \{\, - \frac{1}{4} \,\norm{T_a (F)}_{H_a}^2 \,\}  \nonumber \\ 
							&=& \exp \{\,- \frac{1}{4}\,\int \,( \o^{-1} |\tilde{f}_0|^2 + \o |\tilde{f}_1 |^2 )\, dp \,\}\,.
\eeqa
The GNS representation of $\o_a$, whose triple we denote by $(\p_a, \cH_a, \O_a)$, is the Fock (vacuum) representation of the algebra $\cW(V_a,\s_a)$.

This Fock space construction is possible only for the symplectic space $(V_a, \s_a)$, see e.g. \cite{ams93}, but construction with a canonical non-regular representation exists instead for all the Weyl algebras associated to the symplectic spaces in diagram (\ref{e:def inclusion symplectic}) as illustrated below.

\smallskip
Indicated the extension of $\s_a$ to the space $V_k$ by $\s_k$, for $V_k$ each of the real spaces in (\ref{e:def inclusion symplectic}), the pairs $(V_k, \s_k)$ are real symplectic space, properly included as in the same diagram. 
Only the symplectic spaces $(V_b,\s_b)$ and $(V_e,\s_e)$ are degenerate, with the same degenerate subspace 
\beq\label{e:R}
N:=\{F=\fzu\in V_b :\fzu=0\oplus n,\, n\in \bR \}\cong \bR_d\,,
\eeq
and because of the results on degenerate symplectic forms as in equation (\ref{e: Weyl center}), there exist a common non trivial center of the C*-algebras $\csb$ and $\cse$ given by
\beq\label{e:center of W_a}
\csn=\fZ_b:= \csb\cap\csb'=\cse\cap\cse'=:\fZ_e\,.
\eeq
We introduce now some symplectic isomorphisms to discuss the model in terms of twisted crossed products of Weyl algebras as in Subsection \ref{ss: isom twist}. 
To any function $f_1\in \duS$, we associate three real values 
\beq\label{e:sw grading}
F_+:=\lim_{x\to+\infty} f_1(x)\,, \quad
F_-:=\lim_{x\to-\infty} f_1(x)\,,\quad \Fin:=\frac{1}{2}(F_+ +F_-)\,.
\eeq
Observe that subtracting the real value $F_\infty$ from the second coordinate function $f_1$ of an element $F=\fzu\in V_b, \,V_e$ realizes the quotient maps in equations (\ref{e:quotient S}). 
Hence we  state the following 
\blemma\label{l:psia}
For given an element $F=\fzu$ in $V_b$ (or $V_e$), we denote by $[f_1]:=f_1-\Fin$ the
element in the quotient group $\cS=\duzS/\bR_d$  (or $\duqS=\duS/\bR_d$) as in the quotient maps of equations (\ref{e:quotient S}). Then 
\bdes
\item[i)]
there exist a splitting isomorphism of degenerate symplectic spaces $\psii$ from $(V_b,\s_b)$ to $(V_a,\s_a)\oplus (N,\s_N)$, with $\s_N\equiv 0$, defined as 
\beq\label{e:definition psib}
\barr{rll}
	\psii:& V_b &\longrightarrow  V_a\oplus N=(\dS\oplus\cS)\oplus N \\
		  &F &\longmapsto (f_0\oplus[f_1])\oplus \Fin \,.
\earr
\eeq
The Weyl functor gives an isomorphism of C*-algebras as 
\[
\Psi_b: \csb \longrightarrow \csa\tmax\csn =\csa\tmin\csn\,.
\]
\item[ii)] there exist a splitting isomorphism of degenerate symplectic spaces, from $(V_e, \s_e)$ to $(V_q, \s_q)\oplus (N, \s_N)$ with $\s_N \equiv 0$, also indicated by  $\psii$ and defined by 
\beq\label{e:definition psie}
\barr{rll}
	\psii:& V_e &\longrightarrow  V_q\oplus N=(\dS\oplus\duqS)\oplus N \\
		  &F &\longmapsto (f_0\oplus[f_1])\oplus \Fin \,.
\earr
\eeq
The corresponding isomorphism of C*-algebras given by the Weyl functor is
\[
\Psi_e: \cse \longrightarrow \csq\tmax\csn =\csq\tmin\csn\,.
\]
\edes
\elemma
\prf
$\psii$ is an isomorphism between the real discrete vector spaces $V_b$ and $V_a\oplus N$, so it trivially preserves the symplectic form and the image of the degeneracy subspace $N$ of $V_b$. The last equality about tensor product holds because $\csn$ Abelian. The same for the $V_e$ case. 
\qed
A more general symplectic space isomorphism extending $\psii$ is also possible for the larger symplectic space $(V_f, \s_f)$ itself, once we have picked an element in it. This isomorphism turns out to be essentially unique. 
We begin this construction by defining \emph{the charges} of a generic element $F=\fzu\in V_f$. These are the two real quantities canonically associated to $F$ by 
\beq\label{e:F charges}
F_c := \int_\bR f_0 \, dx,\qquad F_q := \int_\bR \partial f_1 \, dx =  F_+-F_-\,. 
\eeq
A complete useful re-parametrization of the symplectic spaces in diagram (\ref{e:def inclusion symplectic}) is obtained by the choice of a \emph{regularizing element} $T\in V_f$ with non vanishing charges, i.e. such that $T_c\neq 0\neq T_q$.
A useful choice for such an element of $V_f$ is the following: given a function $t\in \duS$ such that 
\beq\label{e:regular element}
\lim_{x\to+\infty} t(x)=-\lim_{x\to-\infty} t(x)=\frac{1}{2}
\eeq
and denoted by $\dt\in \cS$ its derivative, we choice $T=\T\in \cS\oplus\duS$. 
For example, we may take  $t(x)= \frac{1}{\p}\arctan x$, or a half-line or bounded smooth localization obtained from a composition of it with a $C^\infty$ function of compact support. Note that  $\int_\bR \, t(x)\dt(x) \,dx=0$ and the charges associated to $T$ both equate $1$, in fact  
\[
T_c=\int_\bR \, \dt(x)\,dx=\lim_{x\to+\infty} t(x)-\lim_{x\to-\infty} t(x)=T_q=1\,. 
\]
Now, because of the fixing of the regularizing element, we may associate to every element $F\in V_f$ two more real numbers 
\beq
F_n :=\int_\bR  \, f_1 \, \dt \, dx \, , \qquad  \qquad F_r :=\int_\bR  \, f_0 \, t \, dx\,, 
\eeq
and functions
\beq\label{e:secondreg}
\fzdt:=f_0\,-\,F_c \,\dt\in \dS,\qquad
\fut := f_1\,-\,F_q \,t \,-\,\Fin\in \cS\,.
\eeq
In this way, denoting by $(L=N\oplus C, \s_L)$ and $(M=R \oplus Q, \s_M)$ two copies of the elementary symplectic space as in Subsection \ref{ss:Elementary Weyl algebras}, i.e. $N\cong C\cong R\cong Q\cong \bR_d$, and setting
\beq
F_t= \Ft\in V_a\,, \quad F_l = \Fl \in L\,,\quad F_m = \Fm \in M\,,
\eeq
it is possible to state the following
\bprop\label{p: psif}
For every choice of an element $T\in V_f$ as above, for example as in (\ref{e:regular element}):
\bdes
\item[i)]
there exists a non splitting isomorphism of symplectic spaces between $(V_f, \s_f)$ and $(\psi_T(V_f), \s_f)\subsetneq (V_a\oplus L \oplus M, \s_a\oplus\s_L\oplus\s_M)$, defined by
\beq\label{e:definition Rt}
\barr{rl}
\psi_T:& V_f \longrightarrow \psi_T(V_f) \subsetneq V_a\oplus L \oplus M\\
		  &F \longmapsto F_t\oplus F_l\oplus F_m\,.
\earr
\eeq
$\psi_T$ extends the splitting isomorphisms $\psii$ of Lemma \ref{l:psia} for $V_b$ and $V_e$ and reduces to the isomorphisms $(V_k,\s_k)\to (\psi_T(V_k),\s_k)$ of symplectic spaces  when applied to $V_k$, $k=c,q$ respectively, i.e. such that:
\edes
\bitem
	\item $\psi_T$ is a $\bR$-linear continuous bijection on its image, preserving symplectic forms and degeneracy subspaces;
	\item for elements $F=\fzu$ and $G=\gzu$ in $V_f$ we have
\[
\s_f (F, G)= \int \, (f_0 g_1 - f_1 g_0 )\, dx = \s_a(F_t, G_t)  \,+\,\s_L ( F_l, G_l)\, +\, \s_M (F_m ,G_m)\,;
\]
\item 
$\psi_T$ maps $V_c$ into elements with $F_q=0$, $V_q$ into elements with $F_c= F_n=0$ and $V_e$ into elements with $F_c=0$.
\eitem
\bdes
\item[ii)]
the subspace $N\subset V_b$ is invariant under the action of $\psi_T$, i.e. for any $F=(0, \Fin)\in N$ it holds $F\mapsto(0,0)\oplus(0,\Fin)\oplus(0,0)$, and the non trivial center of $\csb$, denoted $\fZ_b$, equals its relative commutant in $\csf$ and $\csc$, i.e. for $k=c, f$ we have 
\[
\fZ_b=\csb^c:=\csb'\cap C^*(V_k,\s_k)\,.
\]
Similarly for $\cse$ instead of $\csb$;
\item[iii)]
for a different choice of  an element $T'\in V_f$ that defines the symplectic isomorphism $\psi_{T'}$, there exists a symplectic isomorphism $\f_{T T'}$ between $\psi_T(V_f)$ and $\psi_{T'}(V_f)$ such that $\psi_{T'}=\f_{T T'}\circ\psi_T $. The isomorphism  $\f_{T T'}$ reduces to the identity on $C\oplus Q$ iff $T-T'\in V_b$, i.e. iff $T$ and $T'$ have the same charge content. Moreover, it exponentiates to a Weyl algebra and C*-algebra isomorphism.
\edes
\eprop  
\prf
The existence of the isomorphisms with the  properties described in i) and ii) follows from easy calculations on the above definitions. The essential uniqueness up to the symplectic isomorphism in iii) follows because $\f_{T T'}$ is defined from the function $T'-T\in V_f$, so that the property of invariance of the charges holds iff $T'-T$ has zero charge content.
\qed
We will refer to the additive group $C\oplus Q$ as the \emph{charge group} of the theory. 
\brem\label{r:isom}
1) $\psi_T$ being  non splitting, its exponentiation $\Psi_T$ gives a Weyl algebra $\bbW(\psi_T(V_f))=\Psi_T(\cW_f)\cong \cW_f$ that is only properly contained in the algebraic tensor product $\cW_a\otimes\cW_L\otimes\cW_M$. The associated C*-algebra is isomorphic to $\csf$ and only properly contained in the respective (maximal) tensor product as well. Hence,  because of the results in Section \ref{s:Weyl}, we have
\beq\label{e:cross weyl prod}
\csf\cong\big((\csa\otimes\fZ_b)\rtimes_{\s_L}\cU(C)\big)\rtimes_{\s_M}\cU(Q)\,.
\eeq
2) Observe that the isomorphism $\psii$ is introduced to quotient out from an element in $V_b$ or $V_e$ its degenerate part in $N$. Instead, the isomorphism $\psi_T$ is tailored to extract from  an element in $V_f$ its charge content, i.e. to fix its equivalence class in the additive group $C\oplus Q$ of the charges, dividing out the remaining components in $V_a$, $N$ and $R$, according to the choice of $T \in V_f$. 
\erem
\subsection{Defining representations for the Streater and Wilde model}
\label{ss: representation sw} 
We define a reference representation for each Weyl and C*-algebra that functorially derives from the six term diagram (\ref{e:def inclusion symplectic}). These extend the Fock representation of the observable algebra and are non-regular representations that satisfy the properties of factorization indicated in Proposition \ref{P:factor state weyl case}. 
The twisted crossed product in (\ref{e:cross weyl prod}), explains the observable and charge content of each of the algebras associated to the symplectic spaces in diagram (\ref{e:def inclusion symplectic}).

\smallskip
We begin by noticing that the equation (\ref{e:cross weyl prod}) gives an interpretation of the non-regular representations introduced for different models of this kind in  \cite{ams93, ams93a}.
In these papers, the representation of $\cW(V_f,\s_f)$ is obtained as the GNS representation associated to the non-regular state
\beq\label{e:ams state}
\o (W_f(F)) = \exp \{- \frac{1}{4} \, q(F) \}\,, \qquad F \in V_f\,,
\eeq
where $q(F)$ is a so called \emph{generalized quadratic form}, defined by
\beq\label{e: g.quadratic}
q(F) = \left\{ \barr{ll}
\int \,( \o^{-1} |\tilde{f}_0|^2 + \o |\tilde{f}_1 |^2 )\, dp \qquad   &\textrm{if}\quad F=\fzu \in V_a \,,\\
+\infty \qquad &\textrm{if}\quad F \notin V_a\,.
\earr
\right.
\eeq
If $\o_a$ is the Fock state on $\cW_a$ and for two symplectic spaces with supports $L\cong M\cong \bR_d^2$ we take two copies $\o_L$ and $\o_M$ of the non-regular elementary state introduced in Subsection \ref{ss:Elementary Weyl algebras}, respectively on the algebras $\cW_L$ and $\cW_M$,  similarly to the equation (\ref{e:ams state}) we give the following
\bdfn\label{d: state d f}
For given $T\in V_f$ a charge regularizing element as in Subsection \ref{ss:WSW model} and notation as in the previous Proposition \ref{p: psif}, and for $\d$ being the Kronecker delta, we define a state on the Weyl algebra $\cW_f$ by
\beqan
\o_f (W_f(F)) &:=& \exp \{ - \frac{1}{4} \,\norm{T_a (F_t)}_{H_a}^2 \}\, 
		\d_{F_c ,0}\,\d_{F_q ,0}\\
&=&\o_a (W_a(F_t))\,\d_{F_c ,0}\,\d_{F_q ,0}\\
&=&\,\o_a (W_a(F_t)) \,\o_L( W_L(F_L)) \,\o_M (W_M(F_M))\,.
\eeqan
\edfn
Observe that the state in Definition \ref{d: state d f} differs from the one in equation (\ref{e:ams state}) only in restriction to elements $W_b(F)\in\cW_b$, with $F\in V_b$ and $F_\infty \neq 0$: the GNS representation of the state $\o_f$, unlike that of $\o$, maps them in elements different from the identity, as seen in Subsection \ref{ss:Elementary Weyl algebras} for the representation $\p_L$. 
This means that $\o_f$, unlike $\o$, results to be faithful on the Weyl elements obtained from the subspace $N$, hence it is preferred in the sequel for obtaining the physical content also for this subspace.

\smallskip
For different regularizing elements, the above definition gives different states, whose GNS representations are actually related by the following 
\bprop\label{p:rappre}
\bdes
\item[i)]
The GNS representations $\p_f$, associated to the states $\o_f$ for different choices $T_1,\,T_2\in V_f$ of the non vanishing charge regularizing element in Definition \ref{d: state d f}, are unitary equivalent. The unitary operator on $\cH_f$ giving this equivalence by adjoint action is $\p_f(W(F))$ for $F=T_1-T_2$; moreover, such a $F\in V_b$ if the charges of $T_1$ and $T_2$ are the same;
\item[ii)]
the representation $\p_f$ is unitary equivalent to the representations $\p_a\otimes \p_L\otimes \p_M\rest\cW_f$ and, denoting by $\cH_{(c,q)}\cong \cH_a$ the Hilbert space at fixed charges $(c, q)\in C \oplus Q$ and $\p_{(c,q)}\cong \p_a$ the relative representation, also to the representation $\oplus_{(c, q)\in C \oplus Q}\, \p_{(c,q)}$. $\p_f$ acts on the Hilbert space
\beq\label{e:0vector}
\cH_f \cong \cH_a \otimes \cH_L \otimes \cH_M \cong \cH_a \otimes \ell ^2 (C)\otimes \ell ^2 (Q)\,\cong \bigoplus_{(c, q)\in C \oplus Q} \cH_{(c,\, q)}\,.
\eeq
Equivalently, we have $\p_b=\p_c\cong \p_a\otimes \p_L$, $\p_q\cong \p_a\otimes \p_M$ and $\p_e\cong\p_f$, where it is meant $\p_b \cong (\p_a\otimes \p_L)\rest_{\cW_b}$ an similar. The representations $\p_a$ and $\p_b$ are the only regular ones; 
\item[iii)]
the GNS vector $\O_f\in \cH_f$ of the state $\o_f$  is the vector
$ |0,0\rangle\in \cH_{(0,0)}\subset \cH_f$. It is cyclic for $\p_f(\cW_f)$ and is identified by the isomorphisms in equation (\ref{e:0vector}) with the vectors 
$ \O_a \otimes \O_L \otimes \O_M$ and $\O_a \otimes \d_0 \otimes  \d_0$ respectively.  Similarly, in the other cases,  $\O_b$ and $\O_e$ not being cyclic for $\cW_b$ and $\cW_e$ respectively.
\edes
\eprop
\prf
i) The equivalence for different $T$ is a consequence of Proposition \ref{p: psif}, because of the isomorphism of the Weyl and C*-algebras and because the states $\o_f$ satisfies the factorization condition (\ref{e:condition 2 weyl}) in Proposition \ref{P:factor state weyl case}. \\ 
ii) It suffices to observe that the state $\o_f$ reduces to the state $\o_a$ on $\cW_a$ and that if $F\in V_f$ with at least a non vanishing charge, then $\o_f(W(F))=0$ because of the Kronecker deltas.\\
iii) is trivial.
\qed
\section{Local theory, DHR Sectors and Gauge Group}\label{s: Local theory SW model rep}
In this section we treat the net approach of AQFT to the Streater and Wilde model. As general reference we use \cite{H,rob00}, and for the low dimensional theories also \cite{klm01}.

\smallskip
Let $M$ be the 1+1-dimensional Minkowski spacetime and consider $\bR$ as its time zero axis.
We recall some notation and fix new  concerning  $\open(M)$ and $\open(\bR)$, the set of open non void subsets of $M$ and $\bR$ respectively, partially ordered by inclusion. 

\smallskip
A causal disjointness relation $\bot$ is induced on $\open(M)$ from the Minkowski metric.
A triple consisting of a subset $\cP\subset\open(\bR)$, the inclusion partial order $\subset$ and the disjointness relation $\bot$ will be called an \emph{index set} and shortly indicated as $(\cP, \subset, \bot)$. 
If $P\in \cP$,  we denote by $P^\bot$ the \emph{causally disjoint set} of $P$, i.e. the set of the elements of $\cP$ disjointed from $P$.

\smallskip
On the Minkowski spacetime the preferred index set is the set of  causally complete bounded connected regions, indicated by $\cK$ and called the set of \emph{open double cones on $M$} and defined as follows:  for given $x,y\in M$ with $y\in V^+_x\subset M$, the \emph{open future cone of $x$} and for $V^-_y\subset M$ the \emph{open past cone} of $y$, a \emph{double cone} $\cO\in \cK$ is defined by the intersection of $V^+_x\cap V_y^-$.

\smallskip
The  relation $\bot$ on $M$ reduces to the set theoretic disjointness relation on the time zero line $\bR$, i.e. for $A,B\in\open(\bR)$ $A\bot B\, \iff \,A\cap B=\emptyset$. 
The index set we work with in this paper, will only be the following:
\beq\label{e:sets}
\cI:=\{ \textrm{non empty, open, bounded intervals of }\bR \}\,.
\eeq
A generic abstract \emph{net $\bbN$ on the index set $\cP$} is defined by an inclusion preserving map
\beq\label{e:gen net}
 \bbN: P \longmapsto \bbN (P)\,,\qquad P \in \cP\,.
\eeq
A net hence takes its image elements in the objects of a generic category, also furnished with the structures of partial order and disjointness.
To distinguish nets defined on different index set and with the same image category, we use the notation $\bbN_\cP$ and define \emph{a maximal element} of the net $\bbN_\cP$ by $\bbN^\cP(M):=\bigvee_{P\in \cP}\bbN_\cP(P)$, where the symbol $\bigvee$ has to be properly understood relative to the category of the image elements. 

\smallskip
A relevant property of a net is \emph{locality}, i.e. $\bbN(P_1)\bot \bbN(P_2)$ for $P_1,\,P_2\in \cP$ and $P_1\bot P_2$; it distinguishes Bosonic (i.e. local) from non Bosonic nets.

\smallskip
For given a symplectic space $(V,\s_V)$ and an appropriate notion of \emph{localization} of an element of $V$ in an element of an index set $\cP$, it remains naturally defined a category of symplectic subspaces with order and disjointness relation $(\sub(V(\cP)),\subset,\bot_{\s_V})$. The partial order is inherited from the one of $\cP$, by the defined localization, and the disjointness of two symplectic subspaces $V_1$ and $V_2$ in this category means that $V_2\in V_1^{\bot_{\s_V}}$, i.e. $\s(v_1,v_2)=0$ for any element $v_1\in V_1$ and $v_2\in V_2$. Hence $(\sub(V(\cP)), \subset,\bot_{\s_V})$ is a first example of the image category of a net that, accordingly  to (\ref{e:gen net}), we indicate by $V_\cP$ and call the net of the symplectic subspaces of the symplectic space  $(V,\s_V)$ on  the index set $(\cP,\bot)$. 
Notice that the notion of localization of the elements of $V$ on $\cP$, also accounts for the locality of $V_\cP$. 

\smallskip
The Weyl functor preserves the net structure on $\cP$, so that a net of Weyl algebra $\cW_\cP: P \longmapsto \cW(V(P), \s_V)$ and a net of C*-algebra $\fN_\cP$ are also defined,  by 
\beq\label{e:nets cstar weyl}
 \fN_\cP :P\longmapsto \fN(P):= \cW \big((V(P),\s)\big)^{-} \,, \qquad P \in \cP\,.
\eeq
Here the completion may be intended with respect to the minimal regular norm.

\smallskip
Once fixed a defining representation $\p_n$, a net $\cN_\cP$ of von Neumann algebras is canonically associated to a Weyl algebra net, by 
\beq\label{e:nets von weyl}
 \cN_\cP :P\longmapsto \cN(P):= \p_n (\cW (V(P),\s))''\,, \qquad P \in \cP\,.
\eeq
If $A\in \open(\bR)$ then $\fN^\cP(A)$ is the  C*-algebra generated by the von Neumann algebras $\cN_\cP(P), \, P \subset A,\, P \in \cP$, so that we may define $\cN^\cP(A):=\fN^\cP(A)''$. Hence, $\fN^\cP(A)$ and $\cN^\cP(A)$ are the C* and von Neumann algebras generated by additivity on the index set $\cP$.
\footnote{
For example, the additivity property for the von Neumann algebras net $\cN_\cP$ means that $\cN(A)=\vee_{P\subset A}\cN_\cP(P)$. 
For the general properties of abstract nets we refer to $\cite{rob00}$, and to $\cite{fabio4}$ for further discussions.
}

In particular, $\fN^\cI(\bR)$ is the C*-algebra generated by all the local von Neumann algebras $\cN_\cI(I)$, for $I \in \cI$. It is called the \emph{C*-algebra of quasi local elements} of the net $\cN_\cI$ and is the C*-algebra referred to in studying the DHR superselection sectors of the net, with respect to the index set $\cI$. The net and this C*-algebra are usually both indicated by the same symbol $\cN$.
\subsection{Nets for the Streater and Wilde model}\label{ss: SW nets}
To define useful nets for the study of the Streater and Wilde model, a proper definition of localization is now in order: for $j=a,\,b,\,c,\,q,\,e,\,f$ we define the localization of a generic element $F=\fzu \in V_j$ as
\beq\label{e:localization}
\loc F:=\supp f_0 \cup \supp \partial f_1\,,\qquad f_0\in \cS,\, f_1\in \duS\,.
\eeq  
According to this definition, the various nets of symplectic subspaces $V_{j,\cP}\subseteq V_{f,\cP}$ with range category  $(\sub(V_j(\cP)),\subset,\bot_{\s_j})$ are defined: explicitly $V_{j,\cI}$ is given by
\[
V_{j\cI}:I\in \cI\longmapsto V_{j\cI}(I):=\{F\in V_j :\loc F\subset I\}\,,\qquad I\in \cI\,.
\]
Observe that the elements $F=0\oplus n\in V_f$, with $n\in N\cong \bR_d$, have vanishing localization Weyl unitaries according to the definition in equation  (\ref{e:localization}), so that they may be thought as localized in any interval of the time zero line.

\smallskip
A possible first result for the model, is a complete \emph{simple current extension} characterization of the nets of von Neumann algebras, derived from the nets of symplectic subspaces in the six term inclusion of symplectic spaces in diagram (\ref{e:def inclusion symplectic}), in the representations defined in Section \ref{s:SW model}. To obtain such a result, we show the existence of \emph{local} symplectic isomorphisms obtained from a natural local choice of the regularizing element $T\in V_f$.  
Such a characterization will be used in Section \ref{ss:gaugess} to discuss the existence of the relative gauge simmetry groups.

\smallskip
The following result collects the local properties of the symplectic isomorphism $\psi_T$,  according to the relative position of $\loc T$ and of the localized algebras. Observe that for $T=\T$ we have $\loc T=\supp \partial t$. We omit for brevity the reference to the symplectic form of the symplectic subspaces involved, and always suppose that the nets are defined on a fixed index set $\cP$.
\bprop\label{p: l t sy i} 
Let $P\subset \bR$ be a generic non void open subset of the time zero real line and $\psi_T$ be a symplectic isomorphism  defined as in Proposition \ref{p: psif} for $T=\T\in V_f(\loc T)$. Letting  $v=F\oplus l\oplus m=(\fzu)\oplus(c\oplus n)\oplus(r\oplus q)$ be the  generic element in $V_a\oplus L \oplus M$, we have
\bdes
\item[i)] if $\loc T \,\bot\, P$ then $\,\psi_T(V_a(P))=V_a(P)\oplus 0\oplus 0$; for a generic $P$ and $\loc T$ it holds
\beqan
\psi_T(V_b(P))&=&\psi_T(V_a(P))\oplus N\,,\\
\psi_T(V_e(P))&=&\psi_T(V_q(P))\oplus N\,;
\eeqan
\item[ii)] 
if $\loc T \subset P$\ then
\beqan
\psi_T(V_c(P))&=&\psi_T(V_b(P))\oplus C=\psi_T(V_a(P))\oplus (C\oplus N)\,,\\
\psi_T(V_q(P))&=&\psi_T (V_a(P))\oplus Q\,,\\
\psi_T(V_e(P))&=&\psi_T(V_b(P))\oplus Q=\psi_T(V_q(P))\oplus N\\\
		&=&\psi_T(V_a(P))\oplus N\oplus Q\,, \\
\psi_T(V_f(P))&=&\psi_T(V_c(P))\oplus Q=\psi_T(V_e(P))\oplus C\\
		&=&\psi_T(V_a(P))\oplus(C\oplus N)\oplus Q\,.
\eeqan
\edes
\eprop
\prf
i) The case $\loc T\, \bot\, P$ is trivial. If $\loc T$ and $P$ are not disjoint, then 
\beqan
\psi_T(V_a(P))&=&\Big\{(f_0,f_1)\oplus \big(0,\mbox{$\int_\bR\, f_1 \,\dt\,dx$}\big)\oplus \big(\mbox{$\int_\bR\, f_0 \,t\, dx $},0\big)\in V_a\oplus 												L\oplus M:\\ 
										&&\quad f_0 \in \dS(P),\, f_1 \in \cS(P)\Big\}
\eeqan
is a proper subset of $V_a\oplus N\oplus R$, that does not split into a symplectic sum. 
Moreover 
\beqan
			\psi_T(V_b(P)) &=& \Big\{ (f_0,f_1)\oplus (0,n)\oplus \big(0,\mbox{$\int_\bR\,f_0 \,t\,dx$}\big)\in V_a\oplus L\oplus M: \\
										 &&\quad f_0 \in \dS(P),\, f_1 \in \cS(P)\Big\}\\
										 &=& \psi_T(V_a(P))\oplus N\,.
\eeqan
The decompositions for $\psi_T(V_q(P))$ and $\psi_T(V_e(P))$ are in the proof of ii) below.\\
ii) The definition of $\psi_T$ as in Proposition \ref{p: psif} for generic $\loc T$ and domain $P$ gives
\beqan
			\psi_T(V_f(P))&=&\Big\{(f_0,f_1)\oplus (c,n)\oplus \big(\mbox{$\int_\bR\,f_0\, t\,dx$},q\big)\in V_a\oplus L\oplus M:\\
										&&\quad f_0\in\dS(P\cup \loc T),\,f_0\,+\,c\,\dt\in \cS(P),\\
										&&\quad f_1\in\cS(P\cup\loc T),\, f_1\,+\,qt\in \duS(P)\Big\}
			\eeqan	
and similarly for the other local symplectic subspaces. Taking $\loc T \subset P$, the results follow.
\qed
Observe that whatever $\loc T$ and $P$ are, with $\loc T \bot P$, the local algebra $\cA(V_a(P))$ defined as in equation (\ref{e:nets von weyl}) satisfies 
\beq
\cA(V_a(P))\cong \p_a\otimes\p_L\otimes\p_M\big(\bbW(\psi_T (V_a(P)))\big)''=\cA(V_a(P))\otimes I_L\otimes I_M\,.
\eeq
Similarly we define from the symplectic local subspaces of $V_b, V_c, V_q, V_e$ and $V_f$ the nets $\cB, \cC, \cQ, \cE$ and $\cF$ respectively. Remembering that we also indicated by  $C$ and $Q$ the additive groups of the charges, we denote without ambiguity by $\cU(C)$ and $\cU(Q)$ their unitary Weyl representations on $\cH_L$ and $\cH_M$ respectively.

Recalling that in Remark \ref{r:isom}, for chosen $T\in V_f$ we denoted by $\Psi_T$ the isomorphism at the Weyl algebra level obtained from the symplectic space isomorphism $\psi_T$ by the Weyl functor, we use the same symbol $\Psi_T$ for its extension to the von Neumann algebras, so that from the representation $\p_f$ to the representation $\p_a\otimes\p_L\otimes\p_M$, it holds $\cA(V_a(P))\otimes I_L\otimes I_M =\Psi_T(\cA(V_a(P)))$, for every $T\in V_f$. 
\footnote{
Such an isomorphism $\Psi_T$ turns out to be weakly continuous only for the closed linear subspaces of the von Neumann algebras with fixed charge, see \cite{fabio4} for detail.
}

Hence, denoting the Abelian algebra of the symplectic subspace $N$ in representation $\p_b$ by  
\beq\label{e:Z_b}
\cZ_b:=\p_b(\cW_b(N))''\,\cong\, \p_L(\cW_L(N))''\,,
\eeq
the above Proposition and the isomorphism $\Psi_T$, give the following characterization of the nets involved, as simple current extensions from the net of observables:
\bprop\label{p:split local lag}
For given $P\subseteq \bR$ a generic non void subset of the line and for notations as above we have 
\bdes
\item[i)] 
for generic $P$ and $\loc T$:
\beqan
\cB(V_b(P))&=&\cA(V_a(P))\otimes\Zb\,,\\ 
\cE(V_e(P))&=&\cQ(V_q(P))\otimes \Zb\,;
\eeqan
\item[ii)] 
if $\loc T \subset I\in \cI$\ then
\edes
\begin{footnotesize}
\beq\label{e:SW general inclusion of net}
\barr{ccccc}
&&\cB(I)&&\cC(I)\\
&&\Vert &&\Vert\\
 \cA(I)& \subset & \cA(I)\otimes\Zb&\subset&(\cA(I)\otimes\Zb)\rtimes_{\s_L}\cU(C) \\\\
 		\cap	&& \cap& & \cap \\\\
		\cA(I)\rtimes_{\s_M}\cU(Q)&\subset&(\cA(I)\otimes\Zb)\rtimes_{\s_M}\cU(Q)&
		\subset&(\cA(I)\otimes\Zb)\rtimes_{\s_M}\cU(Q)\rtimes_{\s_L}\cU(C)\\
\Vert &&\Vert&&\Vert\\
\cQ(I)&&\cE(I)&&\cF(I)\,.
\earr
\eeq
\end{footnotesize}
\eprop
\prf
From the Proposition \ref{p: l t sy i}, by functoriality and the splitting of the representation.
\qed
\subsection{Chiral versus time zero formulation}\label{ss:Chiral versus time zero}
We introduce in the sequel the chiral formulation of the scalar massless free field nets: this allows one to relate the time zero approach to the Streater and Wilde original one and to the conformal chiral one given by Buchholz, Mack and Todorov in \cite{bmt88}.

\smallskip
To fix notation, let $\widetilde{\cA}=\widetilde{\cA}_\cK$ be the observable net of the Streater and Wilde model on the index set of the double cones $\cK$ of the 1+1-dimensional Minkowski spacetime $M$. 

If $I\in \cI$ is a time zero open interval and $\cO=I''\in \cK$ is the open double cone it generates by causal completion, i.e. the interior of the causal closure of the interval $I$, the net $\widetilde{\cA}$ has time zero generating net  $\cA_\cI$, i.e. it holds $\widetilde{\cA}(\cO)=\cA_\cI(I)$.

A wherever based double cone, is described as $\cO=I_+\times I_-$, for $I_\pm \in \cI_\pm$, the index set of the open bounded intervals on the right/left light ray lines,  see e.g. \cite{reh00}.

\smallskip
The Streater and Wilde original model in \cite{sw70} is formulated in the left/right chiral fields formalism and, for every chiral component, it gives a family labeled by $\bR_d$ of DHR automorphisms for the net $\widetilde{\cA}$. Disliking this, we preferred to write the symplectic spaces and nets in a time zero formulation and then, using the symplectic isomorphism $\psi_T$ defined in Proposition \ref{p: psif}, we extracted the charge content. To relate the two approaches, and to discuss the geometric covariance properties, we may use a second symplectic isomorphism constructed as in the sequel. 

\smallskip
In the chiral approach, the symplectic spaces are given by the smearing test functions of the left/right mover solutions $\th_\pm$ of the classical wave equation, i.e. $(\partial_t \pm \partial_x)\th_\pm=0$. 

To consider the charge carrying fields, we have to restrict to the functions $\th_\pm(x)\in C^\infty_\bR(\bR)$  with $\partial \th_\pm \in \cS$ and a general solution of the wave equation can be written as
\footnote{
Actually in Streater's and Wilde's original formulation the unnecessary requirement $\lim_{y\to-\infty}\th_\pm(y)=0$ is made. We omit such a choice that does not reveal the presence of $\Zb$.
} 
\[
\Th(x,t):=\th_-(x-t)+\th_+(x+t)\,.
\]
To give a symplectic isomorphism between the two formulation, we start from a direct application of the d'Alambert formula: if $F=\fzu\in V_f$, we have 
\beqa\label{e:dala}
\Th(x,t)=\frac{1}{2}\Big[\,f_1(x+t)\,+\,f_1(x-t)\,+\,\int^{x+t}_{x-t}\,f_0 (y)\,dy\,\Big]\,.
\eeqa
Denoted by $V :=\{\th \in C^\infty_\bR(\bR): \partial \th\in \cS \}$ a chiral symplectic space support, and by  
\[
\s_\pm(\th,\f)\,=\,\pm\int_\bR (\f\,\partial\th\,-\,\th\, \partial \f)\, dx
\]
two symplectic forms on $V$, we can hence define the \emph{left/right chiral symplectic spaces of the fields} by $V_+:=(V,\s_+)$ and $V_-:=(V,\s_-)$.

\smallskip
Consider now the isomorphism $\psi$ of real linear spaces defined by  
\beq\label{e:dala psie}
\barr{rcl}
\psi:&V_f&\longrightarrow\psi(V_f)\,=\,V\oplus V	\\
	  &F=\fzu &\longmapsto \psi(F):=\Th^F:=\th_+^F\oplus\th_-^F \,,		
\earr
\eeq
where the direct and inverse transformations are explicitly given by
\beq\label{e:dalatrsf}
\begin{aligned}
\left\{ 
\barr{ll}        
\th_+^F(x)=\frac{1}{2}\{f_1(x)+\int^x_{-\infty}f_0(y)\,dy\}\\
\th_-^F(x)=\frac{1}{2}\{f_1(x)-\int^x_{-\infty}f_0(y)\,dy\}
\earr
\right.
\end{aligned}
,\quad
\begin{aligned}         
\left\{
\barr{ll}
f_0(x)=\partial(\th_+^F(x)-\th_-^F(x))\\
f_1(x)=\th_+^F(x)+\th_-^F(x)
\earr
\right.
\end{aligned}
.
\eeq
Define moreover for every $\Th =(\th_+,\th_-)\in V\oplus V$ the two real constants 
\[
\Th(+\infty):=(\th_+(+\infty),\th_-(+\infty))\in \bR_d^2
\]
where
\beq\label{e:th non F infnity}
\th_\pm(+\infty):=\lim_{x\to +\infty}\th_\pm(x)\,.
\eeq
Observe that for $F\in V_f$ and $\Th=\psi (F)$, this pair of constants is given by 
\beq\label{e:th infnity}
\th^F_\pm(+\infty) = 
	\lim_{x\to +\infty}\frac{1}{2}\Big[f_1(x)\pm\int^x_{-\infty}f_0(y)\,dy\Big]\,.
\eeq
Introducing the (continuous) real valued $\bR$-linear form $\a:\bR^2\to \bR$ such that $\a(a, b)= ab$, we may denote by $\s_\infty$ the usual symplectic form on $\bR^2$, defined by 
\beq\label{e:simp infyn}
\s_\infty ((a_1, b_1), (a_2, b_2))= \a(a_1 b_2) - \a(b_1 a_2)=a_1 b_2 - b_1 a_2\,,
\eeq
and finally state the following
\blemma\label{l:chiral}
There exists a non splitting symplectic isomorphism $\psi_\infty$, equating the isomorphism $\psi $ of equation (\ref{e:dala psie}) as a real linear space isomorphism, defined by
\beq\label{e:dalla psie}
\barr{rll}
\psi_\infty:&(V_f,\s_f)&\longrightarrow\psi_\infty((V_f,\s_f))= (V_+\oplus V_-,\s_+ + \s_- +\s_\infty)\\\\
	  &F=\fzu &\longmapsto \Th^F:=(\th^F_-,\th^F_-)\,,  		
\earr
\eeq
such that for the symplectic forms it holds
\[
\s_f(F,G)\,=\,\s_+(\th^F_+,\th^G_+)\,+\,\s_-(\th^F_-,\th^G_-)\,+\,\s_\infty(\Th^F(+\infty),\Th^G(+\infty)),\quad F,G\in V_f\,.
\]
The corresponding Weyl algebras isomorphism is 
\beq\label{e:twist infty}
\Psi_\infty: \cW(V_f, \s_f)\to \Psi_\infty(\cW(V_f, \s_f))= 
\cW(V_+,\s_+)\rtimes_{z_-}\cU(V_-)
\eeq
where in the $2$-cocycle $z_-=(\b, y)$, the action $\b$ is defined by the $\a$ appearing in equation (\ref{e:simp infyn}) and the $\bT$-valued function $y$ is defined by the symplectic form $\s_-$.   
\elemma
\prf
The proof it trivial, it is enough to reefer to the general case treated in Subsection \ref{ss: isom twist}, in particular to the equations (\ref{e:azione prodotto}) and (\ref{e:cociclo prodotto}) for the definition of $z_-$. Observe that the part $\s_\infty$ in the symplectic form, corresponds to the interacting part of $\s_f$ (actually denoted by $\s_{V_+,V_-}$ according to the notation of Subsection \ref{ss: isom twist}).
\qed
Various observations are now in order:
\bitem
\item 
\emph{Picking a point at infinity} $\infty\in \bR\cup\{\pm\infty\}$ in the limits of the integrals appearing in equations (\ref{e:dalatrsf}) is necessary in order to define $\psi_\infty$ and $\s_\infty$. 
The choice $\infty =-\infty$ gives  $\th_-^F(-\infty)=\th_+^F(-\infty)=\frac{1}{2}f_1(-\infty)$, so that the \emph{interacting part} $\s_\infty$ of the symplectic form depends only on the limit value of the functions at $+\infty$.
\item
The space $\psi_\infty(V_f)$ is not a direct sum because $\psi_\infty$ is not splitting. As a consequence, the Weyl algebra $\Psi_\infty(\cW(V_f, \s_f))$ is not a tensor product but a twisted crossed product of Weyl algebras. Changing the role of $V_+$ and $V_-$ in the equation (\ref{e:twist infty}) we also have $\cW(V_f, \s_f)\cong \cW(V_-,\s_-)\rtimes_{z_+}\cU(V_+)$, where now the $2$-cocycle $z_+$ is given in terms of $\a$ and $\s_+$. To make evident the symmetry behind this construction, and its physical significance, we use the following notation 
\beq\label{e: rinfty}
\cW(V_f, \s_f) \,\cong \,\cW(V_+,\s_+)\,\timeinfty\, \cW(V_-,\s_-)\,,
\eeq
where the symbol $\timeinfty$ accounts for the \emph{twisted interaction} operated  by $\s_\infty$ between the two chiral field algebras at the chosen point $\infty$. 
\item
For $\cS$ the Schwartz space of functions on the chiral line and $V_a\cong \cS$
we define the two \emph{chiral symplectic space of observables} as $V_{a_\pm}:=(V_a,\s_\pm)\subset (V_\pm, \s_\pm)$. 
Choosing $\infty =-\infty$ we have $\Th(+\infty)=(0,0)\in \bR_d$, for any $\Th\in V_{a+}\oplus V_{a-}$. Hence  $\s_\infty$ vanishes, independently of the choice of $\infty$ we made, when at least one of its arguments is in $V_{a_+}$ or $V_{a_-}$.
We may interpret this as and independence relation, given by the vanishing of the commutator at $\infty$ between any field of one chirality and any observable of the other chirality.
\item
The isomorphism  $\psi_\infty$ splits when restricted to the observable symplectic subspace, i.e. it hold $\psi_\infty(V_a,\s_a)=(V_{a_+},\s_+)\oplus(V_{a_-},\s_-)$, and we have
\[
\cW(V_a,\s_a)\,\cong\,\cW(V_{a_+},\s_+)\otimes \cW(V_{a_-},\s_-)\,.
\]
\eitem
The (vacuum) states of  the algebras $\cW(V_{a_\pm},\s_\pm)$, i.e. the quasi-free state on the chiral observables algebras $\cW(V_{a_\pm},\s_\pm)$, are defined for $\th \in V_{a_\pm}$ and $T_{a_\pm} (\th)(p):=|p|^{\frac{1}{2}}\,\tilde{\th}(p)$ by
\beq\label{e:Fock chiral state}
\o_{a_\pm} (W(\th)) := \exp \{\, - \frac{1}{2} \,\norm{T_{a_\pm} (\th)}_{H_{a_\pm}}^2 \,\} 
      = \exp \{\,- \frac{1}{2}\,\int_\bR \, |p| \,|\tilde{\th}|^2 \, dp \,\}\,.
\eeq
Similarly to the time zero formulation, the GNS construction from these states gives Fock space representation on the Hilbert spaces $\cH_{a_\pm}$,  with chiral one particle spaces $H_{a_\pm}:=T_{a_\pm} (V_{a_\pm})^-$, and  GNS (vacuum) vector $\O_{a_\pm}$, see for example \cite{bmt88} for details.
From these representations of the chiral observable algebras, we can obtain the non-regular representations of the chiral field algebras, using the regularizing elements tool of Subsection \ref{ss:WSW model} as follows. 

\smallskip
Initially we observe that a simple relation between the charges carried by a field in the two formulations is also obtained from the d'Alambert formula (\ref{e:dala}). The charges carried by $\Th=(\th_+,\th_-)\in V_+\oplus V_-$ are the pair $(c_+,c_-)\in C_+ \oplus C_-\cong \bR_d \oplus \bR_d$ where
\beq\label{e:charges SW}
c_\pm :=\lim_{y\to+\infty}\th_\pm(y)- \lim_{y\to-\infty}\th_\pm(y)\,.
\eeq 
If $F\in V_f$ and if $c_\pm$ are the charges of $\psi_\infty (F)=\Th^F$, we have 
\beq\label{e:cariche trasf}
F_c=c_+ -c_- \qquad \textrm{and}\qquad F_q=c_+ + c_-\,. 
\eeq
In particular we have that $(F_c, F_q)=(0,0)$ iff $(c_+,c_-)=(0,0)$.

\smallskip
Consider now two copies of the elementary symplectic space, defined in Subsection \ref{ss:Elementary Weyl algebras}, that we denote by $(L_\pm, \s_L)$, with $L_\pm= C_\pm\oplus N_\pm\cong \bR_d^2$.
Let moreover $\o_{L_\pm}$ be the non-regular states on the Weyl algebras $\cW(L_\pm,\s_{L})$ respectively, defined as in equation (\ref{e:state NT}) by $\o_{L_\pm}(W((c_\pm,n_\pm))=\d_{c_\pm,0}$, for $(c_\pm,n_\pm)\in L_\pm$ and $\d$ the Kronecker delta.

\smallskip
A regularizing element for the left field movers (similarly for the right movers case), is a non vanishing charge element $S_+$, i.e. $S\in V_+ / V_{a_+}$. We can choose it  such that $\int_\bR S_+ \, \partial S_+ dx=0$ for simplicity.  
Then, as in Proposition \ref{p: psif}, there exist a regularizing isomorphism 
\beq\label{e:regchir}
\barr{rcl}
\psi_{S_+}:&(V_+,\s_+)&\longrightarrow\psi_{S_+}((V_+,\s_+))\subsetneq(V_{a+},\s_+)\oplus(L_+,\s_{L}) \\\\
	  &\th &\longmapsto \th -c_+ S_+\oplus (c_+,n_+)
\earr
\eeq
where the charge $c_+=\lim_{y\to+\infty}\th(y)- \lim_{y\to-\infty}\th(y)$ and $n_+:=\int_\bR \,(\th \partial S_+ - S_+ \partial \th)\, dx$. Observe that $c_+=0$ if $\th\in V_{a_+}\subset V_+$. Similarly, we obtain $(c_-,n_-)\in L_-$ and, as in Proposition \ref{P:factor state weyl case}, it is easy to see that two non-regular states $\o_\pm$ on $\cW(V_\pm,\s_\pm)$ respectively, are defined by 
\beq\label{e:omega plus}
\o_\pm (W(\th))=\o_{a_\pm}(W(\th -c_\pm S_\pm))\,\o_{L_\pm}((c_\pm, n_\pm))\,,\qquad\qquad \th \in V_\pm\,, 
\eeq
with non separable GNS representation spaces $\cH_\pm\cong \cH_{a_\pm}\otimes \cH_{L_\pm}$ respectively.

\smallskip
A state for the algebra $\cW(V_+,\s_+) \,\timeinfty \,\cW(V_-,\s_-)$ is defined by 
\beq\label{e:state chiral}
\o_\infty (W(\Th))\,:=\,\o_{a_+}(W(\th_+ -c_+S_+))\,\o_{a_-}(W(\th_- -c_-S_-))\,\d_{c_+,0} \,\d_{c_-,0}\,,
\eeq
for the element $W(\Th)\in \cW(V_+,\s_+)\, \timeinfty\, \cW(V_+,\s_+)$ associated to the pair $\Th=\th_+ \oplus\th_- \in V_+\oplus V_-$, with charges 
$(c_+, c_-)\in C_+ \oplus C_-$. 
Observe that the definition of the state $\o_\infty$ above is well posed, in fact once we write the obvious algebraic isomorphism  
\[
\cW(V_+,\s_+)\, \timeinfty\, \cW(V_-,\s_-)\cong\cW(V_{a_+}\oplus N_+,\s_+) \otimes \cW(V_{a_-}\oplus N_-,\s_-)\rtimes\cU(C_+ \oplus C_-) \,,
\]
the condition (\ref{e:condition 2 weyl}) of Proposition  \ref{P:factor state weyl case} is satisfied because for the state 
$\o_C = \o_{L_+\oplus L_-}\rest_{C_+ \oplus C_-}$ it holds $\o_C(W(c_+,c_-))=\d_{(0,0),(c_+,c_-)}$.

\smallskip
Moreover, the choice of the point at infinity $\infty$ (note that the values of the functions at the point at infinity are not involved in the definition of $\o_\infty$) and of the regularizing elements $S_\pm \in V_\pm$, defines the state $\o_\infty$ uniquely, up to isomorphism, as in point i) of Proposition \ref{p:rappre}.

\smallskip
From the GNS representations $\p_\pm:=(\p_{a_\pm}\otimes \p_{L_\pm})\circ\psi_{S_\pm}$ of the Weyl algebras $\cW(V_\pm, \s_\pm)$ respectively, we obtain for the non-regular GNS representation of the state $\o_\infty$ :
\[
\p_\infty\cong \p_+\otimes\p_-:=((\p_{a_+}\otimes \p_{L_+})\circ \psi_{S_+})\otimes ((\p_{a_-}\otimes \p_{L_-})\circ \psi_{S_-})\,.
\]
For the Hilbert spaces $\cH_\infty$ of this representation it holds $\cH_\infty\cong\cH_+\otimes \cH_-\cong \cH_f$, i.e. $\p_\infty$ is unitarily equivalent to the representation $\p_f$. In fact, for $F=(\fzu)\in V_a$ we have
\beqa
\norm{T_a (F)}_{H_a}^2 \, 
      &=&\, \int_\bR \,( |p|^{-1} \,|\tilde{f_0}|^2 +|p|\,|\tilde{f_1}|^2)\, dp 
      \,=\, \int_\bR\, 2\,\,|p|\,|\widetilde{\th^F_+}|^2\,+\, 2\,|p|\,|\widetilde{\th^F_-}|^2\, dp   \nonumber\\  				
      	&=&\,2\,\norm{T_{a_+} (\th^F_+)}_{H_{a_+}}^2 \,+\, 
     					2\,\norm{T_{a_-} (\th^F_-)}_{H_{a_-}}^2\,.
\eeqa
We pass now to define the chiral and 1+1-dimensional nets of the fields and their relation with the time zero one. 

\smallskip
For a time zero based double cone $\cO=I''=I_+\times I_-$, with $I \in \cI$ and $I_\pm\in \cI_\pm$, we have:
\beq\label{e:chiraltensfld}
\widetilde{\cF}(\cO)\,=\,\cF(I)\,\cong\,\cF_+(I_+)\,\timeinfty\, \cF_-(I_-)\,.
\eeq
Here $\timeinfty$ denotes the twisted crossed product for the point at infinity obtained in the above equation (\ref{e: rinfty}) and  $\cF_\pm:I_\pm \mapsto \p_\pm(\cW(V_\pm(I_\pm))''$ are the chiral field nets on  the non separable Hilbert spaces $\cH_{_\pm}$ defined, for $I_\pm\in \cI_\pm$ the light ray intervals, from the localized symplectic space $V_\pm (I_\pm)= \{\th\in V_\pm: \supp\partial V\subset I_\pm\}$. 

\smallskip
The observable chiral nets are defined on the separable Hilbert Fock spaces $\cH_{a_\pm}$ by  
\[
\cA_\pm:=I_\pm \mapsto \p_{a_\pm}(\cW(V_{a_\pm}(I_\pm))''\,, \qquad \qquad I_\pm\in \cI_\pm\,,
\]
and the following equality holds for $\cO$ as above 
\beq\label{e:chiraltensobs}
\widetilde{\cA}(\cO)\,=\,\cA(I)\,\cong\,\cA_+(I_+)\otimes\cA_-(I_-)\,.
\eeq
As a last observation, notice that if $N=\{0\oplus n,\, n \in \bR\}\subset V_f$ is the time zero subspace defined in above Subsection \ref{ss:WSW model}, than $\psi_\infty(N) =N\oplus N \subset V_+\oplus V_-$, because $\psi_\infty(n)=\frac{n}{2}$. Hence $\psi_\infty(N)$ generates the diagonal constant elements in the field, and it holds
\[
\cZ_b\cong\p_+(\cW(\psi_\infty(N)))''\otimes\p_-(\cW(\psi_\infty(N)))''
\cong \cap_{I_\pm\in \cI_\pm}\cF_\pm (I_\pm)\,,
\]
i.e. the common Abelian von Neumann subalgebra of the local left and right movers is isomorphic to the non trivial center $\cZ_b$ of the time zero net $\cB$.

\smallskip 
It is well known that the 1+1-dimensional observable net $\widetilde{\cA}$, in the vacuum representation $\p_a\cong \p_{a_+}\otimes\p_{a_-}$, satisfies (see \cite{sw70,hl82,bgl93} for this model and \cite{kl04} for a general formulation): 
\bitem
\item
covariance under the action of $\ov{PSL}(2,\bR)\times\ov{PSL}(2,\bR)$, the universal covering group of the M\"{o}bius group on $M$, implemented by a continuous, unitary, positive energy representation that we denote by $U_a=U_{a_+}\otimes U_{a_-}$; 
\item
Reeh-Schlieder property for $\widetilde{\cA}$ relative to the cyclic and separating vector $\O_a\in \cH_a$ (and also for the net $\cA$, relative to the same vector, and for $\cA_\pm$ relative to $\O_{a_\pm}\in \cH_{a_\pm}$); and
\item
the local algebras $\widetilde{\cA}(\cO)$ ($\cA_\cI(I)$ and $\cA_\pm(I_\pm)$) appear as type $III_1$ factors. 
\eitem
Moreover, as a consequence of the modular theory for the local algebras and of the M\"{o}bius covariance, it is also shown in \cite{hl82} that we have 
\bitem
\item
Haag duality for the net $\widetilde{\cA}$, with respect to  the double cone index set $\cK$ (respectively of $\cA_\cI$ w.r.t. the index set $\cI$ and nets $\cA_\pm$ w.r.t. the index sets $\cI_\pm$); 
\item
timelike duality for the net $\widetilde{\cA}$, w.r.t. the index set $\cK$, and
\item
unitary equivalence of the local algebras $\widetilde{\cA}(\cO)$ with any local algebras defined by additivity on bounded simply connected regions or wedges in $M$ (respectively of the algebra $\cA_\cI(I)$ and the algebra defined by additivity on the half lines, in time zero formulation).
\eitem
\subsection{DHR sectors for the observable net $\cA_\cI$}
\label{ss:dhr gauge group}
We begin this section recalling some basic facts about DHR superselection theory, see e.g. \cite{rob00}.

\smallskip
Given a directed index set $\cP$ and a von Neumann algebras net $\cN_\cP$ in representation $\p_n$, it holds $\cN_\cP\subset \cN^\cP:=(\p_n,\p_n)'$. The DHR superselection sectors of $\cN_\cP$ are described by the $W^*$-category $\repb \cN_\cP$ defined as follows:

\bitem
\item
objects: the representations $\p$ of $\cN_\cP$ such that 
\bitem
\item
$\p$ is a  representation \emph{in} $\cN^\cP$, i.e. $\p( \cN_\cP(P))\subset \cN^\cP$ for $P\in \cP$;   
\item
$\p$ satisfies the DHR \emph{superselection criterion}, i.e.
\beq\label{e:DHR crit}
\p\rest P^\bot\,\cong\,\p_n \rest P^\bot\,, \qquad P\in \cP\,;\qquad \textrm{and} 
\eeq
\eitem
\item
arrows: the intertwiners between these representations, i.e. operators $T\in \cN^\cP$ such that for any $P\in \cP$ and $\p_1,\p_2 \in \repb \cN_\cP$ we have 
\beq\label{e:intretww}
T\p_1(A)=\p_2(A)T\,, \qquad A\in \cN_\cP(P)\,. 
\eeq
\eitem
By \cite{dhr69I, dhr71} it is possible to analyze $\repb \cN_\cP$ in terms of the tensor $W^*$-category $\cT_t$ of \emph{localized transportable endomorphisms} of the net $\cN_\cP$ (if $\cP$ is directed this is actually an equivalence of $W^*$-categories). 
The correspondence functor is given on the objects of $\cT_t$ by $\p=\p_n\circ\r$, for $\r\in \cT_t$ and $\p\in \repb \cN_\cP$. 
\footnote{
The endomorphism $\r$ is said to be \emph{localized} in $P\in \cP$, where $P$ is the element appearing in equation (\ref{e:DHR crit}); moreover it is said to be \emph{transportable} if there exists a field of endomorphisms $\cP\ni a\mapsto\r^a$ such that $\r^a=\r$ if $a=P$, the localization region of $\r$ and if for every $b\in \cP$ such that $a, P\subset b$, there exists a unitary operator $u(b)\in \cU(\cH_\p)$ such that $\r^a=\ad u(b)\circ \r$. 
For a general approach, also on non directed set, see \cite{rob00}.
}
This functor allows one to reduce the study of all the representations to the Hilbert space $\cH_n$, with intertwiners defined as in equation (\ref{e:intretww}). 

\smallskip
In our case, using the above recalled identification of $\repb \cA_\cI$ and $\cT_t$, we define for any $F\in V_f(I)$ that carries the charges $(F_c, F_q)\in \dG$ a DHR representation for the nets $\cA_\cI$ and $\widetilde{\cA}$, by the adjoint action automorphism $\r^I_F$ obtained by $F$ and localized in $I$, such that 
\beq\label{e:sector autom}
\p_a\circ\ad \p_f (W_f (F)):=\p_a \circ \r^I_F\,.
\eeq
For the representations in $\repb \cN_\cP$ the covariance property under the action of a geometrical symmetry group of the spacetime is a standard requirement. 

\smallskip
In our model, in order to define and discuss such a covariance and the positivity of the energy for the net $\cA_\cI$ and for the introduced subsidiary nets, we have to pass to the 1+1-dimensional theory. In fact it is well known that if $\widetilde{\cN}_\cK$ is a (Poincar\'e or M\"{obius}
\footnote{
The name \emph{conformal covariance} is better reserved for covariance under the diffeomorphisms group, that will not appear in this paper.
})
translation covariant net on the index set $\cK$ of the double cones in a 1+1-dimensional Minkowski spacetime $M$, its time zero restriction net $\cN_\cI$ is only a space-translation covariant net, without spectrum condition, see for example \cite{klm01}. 

\smallskip
A representation $\p$ of a net $\cN_\cK$ is said to be \emph{M\"{o}bius  covariant} if there exist a unitary representation $U_\p$ of the group $\ov{PSL}(2,\bR)\times\ov{PSL}(2,\bR)$ on the Hilbert space $\cH_\p$ such that, for any double cone $\cO=I_+\times I_-$ and element $g=(g_+, g_-) \in \cU\subset \ov{PSL}(2,\bR)\times\ov{PSL}(2,\bR)$ in the connected neighborhood $\cU$ of the identity element of the covering of the M\"{o}bius group, we have
\beq\label{e:moeb cov}
\ad U_\p (g)\big(\p(\cN(I_+\times I_-))\big)\,=\,\p(\cN( g_+ I_+\times g_- I_-))\,.
\eeq
Returning to the Streater and Wilde model, if we take $F\in V_f(I)$  implementing the sector automorphism $\r^I_F$ and the representation $\p^I_F:=\p_a\circ \r^I_F$ as in equation (\ref{e:sector autom}) labeled by the charges $(F_c, F_q)\in \dG$, we may pass to the chiral formulation using the symplectic isomorphism $\psi_\infty$ of Lemma \ref{l:chiral}. For given the chiral charges $(c_+^F,c_-^F)=(\frac{F_q+F_c}{2},\frac{F_q-F_c}{2})$, see equation (\ref{e:cariche trasf}), we have a unitary representation of $\ov{PSL}(2,\bR)\times\ov{PSL}(2,\bR)$ on the subspace 
\[
\cH_{c_+^F}\otimes\cH_{c_-^F}\subset \cH_+\otimes \cH_-\,\cong\,(\oplus_{c_+\in C_+}\,\cH_{c_+})\otimes (\oplus_{c_-\in C_-}\,\cH_{c_-})\,. 
\]
For every chiral charge $c_\pm\in C_\pm$, and for $\cH_{a_\pm}$ the observables separable Hilbert space, we have the Hilbert space equivalence $\cH_{c_\pm}\cong \cH_{a_\pm}$, but carrying different representations. If this equivalence is given by the unitaries $Y_\pm: \cH_{a_\pm}\to \cH_{c_\pm}$ and if $U_{a_\pm}$ are the representations of the M\"{o}bius  group on $\cH_{a_\pm}$ respectively, then we may define by $U_{c_\pm}:=\ad Y_\pm(U_{a_\pm})$ the unitary representations of the M\"{o}bius  group on the Hilbert spaces $\cH_{c_\pm}$. The representations of the same group on $\cH_{c_+}\otimes \cH_{c_-}$ are defined by $U_{c_+, c_-}:=U_{c_+}\otimes U_{c_-}$.

\smallskip
Denote now by $\xi_\pm \in I_\pm$ the coordinates on ray light lines, and take an element $\Th=(\th_+,\th_-)\in V_+(I_+)\oplus V_-(I_-)$, i.e. with $\supp \Th =I_+\times I_-= \cO$, and charges equal to $(c_+,c_-)$. For a given DHR representation $\p_{c_+}\otimes \p_{c_-}$, and for any $g=(g_+, g_-) \in \cU\subset \ov{PSL}(2,\bR)\times\ov{PSL}(2,\bR)$, the symmetry group representation $U_{c_+, c_-}$ acts as
\beqa\label{e:covariance reps}
&&\ad U_{c_+,c_-} (g)\,\big(\p_{c_+}(W(\th_+(\xi_+)))\otimes \p_{c_-}(W(\th_-(\xi_-)))\big) \nonumber \\
&&\qquad \quad=\ \ad U_{c_+}(g_+)\,\big(\p_{c_+}(W(\th_+(\xi_+)))\big)\,\otimes\, \ad U_{c_-} (g_-)\, \big(\p_{c_-}(W(\th_-(\xi_-)))\big)\nonumber \\
&& \qquad \quad =\ \p_{c_+}\big(W(\th_+(g_+^{-1}\xi_+))\big)\,\otimes\, \p_{c_-}\big(W(\th_-(g_-^{-1}\xi_-))\big)\,.
\eeqa
Observe that it is not necessary to define the action of the M\"{o}bius group on the two left/right chirality points at infinity $\infty_\pm$, that not even explicitly appear in the action of the representation $U_{c_+,c_-}$ for no $(c_+,c_-)\in C_+\oplus C_-$. In fact it is possible to choose $\infty_+$ not contained in the left supports $I_+$ of $\Th$, and take the open connected neighborhood $\cU$ such to remain with $\infty_+ \notin g_+ I_+$, for every $(g_+,g_-)\in \cU$, and similarly for $\infty_-$.

Moreover, the defined representation $U_{c_+,c_-}$ may also be used in restriction to the elements of the chiral observable net $\widetilde{\cA}_\cK=\cA_+\otimes\cA_-$ in any representation $\p^I_F\cong \p_{c_+}\otimes\p_{c_-}$. 

Finally, the covariance of the vacuum representation $\p_\infty\cong \p_+\otimes\p_-$ of the field net $\widetilde{\cF}_\cK$ is also easily obtained, defining the M\"{o}bius group representation by
\beq\label{e:mgroup f}
U:=\oplus_{(c_+,c_-)\in C_+\oplus C_-}\,U_{c_+,c_-}=(\oplus_{c_+\in C_+}\,U_{c_+})\,\otimes \,(\oplus_{c_-\in C_-}\,U_{c_-})\,,
\eeq
under the proper notion of convergence, i.e. summing up on finite sets of charges $\L\subset C_+ \oplus C_-$.
The same result holds for the time zero formulation, under the isomorphisms seen above. 

\smallskip
After this excursion on the chiral formulation and the representation of the M\"{o}bius group, the reason why we choose time zero description is going to be clear in a moment.

\smallskip
\emph{Solitonic sectors} appeared as part of AQFT in \cite{rob76}, for a general formulation see also \cite{fro76,reh97}. 
In the Streater and Wilde model this notion emerge for the sectors of the intermediate net $\cC_\cI$, that turn out to be similar to a class studied in \cite{mug99} for massive and conformal field theories. 

\smallskip
For $I\in \cI$, we define its \emph{left and right causally disjoint sets} $I^\bot_l,\,I^\bot_r\subset \cI$, by $I^\bot_l\cup I^\bot_r = I^\bot$ and $I^\bot_l\cap I^\bot_r = \emptyset$, and use the following 
\bdfn\label{d:soliton}
Let $\cN_\cI$ be a net in the defining representation $\p_n$, and let $K$ be a group with a local action $\a$ on the net $\cN_\cI$. A translation covariant representation $\p$ of the net $\cN_\cI$ is said to be a \emph{$K$-solitonic representation with support $I\in \cI$} if there exist $h,k\in K$ such that for any $I_1\in I^\bot_l$ and $I_2\in I^\bot_r$ we have
\[
\p\rest\cN(I_1)=\p_n\circ \a_h (\cN(I_1))\,,\qquad \qquad  \p\rest\cN(I_2)=\p_n\circ \a_k (\cN(I_2))\,.
\] 
A \emph{$K$-solitonic automorphism}  $\r_{k,h}^I$ of the net $\cN_\cI$ is a net automorphism such that
$\p=\p_n\circ\r_{k,h}^I$ is a solitonic representation with support $I\in \cI$, i.e.   $\r_{k,h}^I\rest\cN_\cI(I_1)=\a_k$ and $\r_{k,h}^I\rest\cN_\cI(I_2)=\a_h$.
\edfn
We remember other notions, presented e.g. in  \cite{mug01,mug04}, also useful for the case at hand. 
 
If $K$ is a (finite) compact group and $\cR_\cI:=\cN_\cI^K$ is the fixed-point net of $\cN_\cI$ under the action of $K$ so that the DHR sectors of the net $\cR_\cI$ are described by $K$-solitonic automorphisms of net $\cN_\cI$, we speak of an \emph{orbifold model}; moreover, such a model is said to be \emph{holomorphic} if the net $\cN_\cI$ has only trivial superselection sectors. 
Starting from a 1+1-dimensional net $\widetilde{\cN}_\cK$, with time zero restriction $\cN_\cI$, being the solitonic automorphisms not locally normal at infinity, it is only as representations of the fixed-point subnet $\widetilde{\cR}=\widetilde{\cN}_\cK^K$, with time zero restriction $\cR_\cI$, that they represent true positive energy DHR sectors. 

In the Streater and Wilde model the properties  of the sector automorphisms of the observable net $\cA_\cI$ (and $\widetilde{\cA}$) implemented by the Weyl elements of the auxiliary nets, are collected in the following result, where geometric covariance properties are also discussed  
\bprop\label{p:sectors autom}
In the above notation, we have
\bdes
\item [i)] for $F=\fzu \in V_e (I)$, the adjoint action of the represented Weyl element $\p_f(W_f (F))$ implements an $N$-solitonic transportable automorphism $\r^I_F$ of the net $\cC_\cI$, localized in $I\in \cI$, i.e. such that  for $G \in V_c$ we have
\beq\label{e:solitonsw}
 \r^I_F( \p_f (W_f (G))):=\ad \p_f (W_f (F)) (\p_f (W_f(G)))= e^{i \s_f(F, G)}\p_f (W_f (G))\,.
\eeq
Moreover, if $ \lim_{x\to \pm\infty} f_1(x)=F_\pm\in N$ and if $\,\loc G \in I^\bot_r$,  we have
\[
\r^I_F \p_f ((W_f (G)))= \a_{F_+} ((W_f (G))))=e^{-i F_+ G_c }\p_f (W_f (G))\,,
\] 
or if $\,\loc G \in I^\bot_l$ we have
\[
\r^I_F \p_f ((W_f (G)))= \a_{F_-} ((W_f (G))))=e^{-i F_- G_c }\p_f (W_f (G))\,.
\]
The elements in $\cC_\cI$ intertwine solitonic automorphisms with the same values of $F_q=F_+-F_-$, preserving supports, if localized with the same support as the automorphisms.
Such automorphisms turn out to be DHR sector automorphisms in restriction to the nets $\cB_\cI$ and $\cA_\cI$, localized in  $I$, with charge $F_q\in Q$ and intertwiners in $\cB_\cI$ and $\cA_\cI$ respectively;
\item[ii)] 
for $F\in V_c(I)$ the automorphism  $\r^I_F:=\ad \p_f(W_f (F))$ is a DHR automorphism of the net $\cE_\cI$, localized in  $I$, i.e. such that for $G \in V_e$ equality (\ref{e:solitonsw}) holds and $\r^I_F\rest \cE(I_1)=\i$ if $I_1\bot I$. The elements in $\cE_\cI$ intertwine automorphisms with same charge, preserving supports if localized as the automorphisms are. The restriction to the nets $\cQ_\cI$ and $\cA_\cI$ gives DHR sector automorphisms for these nets as well, with intertwiners in $\cQ_\cI$ and $\cA_\cI$ respectively;
\item[iii)]
for any $F\in V_f(I)$ the automorphism $\r^I_F$ defined as in equality (\ref{e:solitonsw}) is equivalent to a positive energy, M\"{o}bius covariant DHR sector automorphism $\r^\cO_F$ for the 1+1-dimensional net $\widetilde{\cA}$, with charges $(F_c,\, F_q)\in \dG$, localized in $\cO=I''$. The intertwiners of two such automorphisms are given by elements in $\widetilde{\cA}$.
\edes
\eprop
\prf
i) The transportability is a result of a simple calculation, see for example \cite{sw70}, and using the correspondence of the given automorphism sectors on the time zero axes. Moreover $\p:=\p_c\circ\r^I_F$ satisfies Definition \ref{d:soliton} for the group $N$, since
\[
F_\pm \mapsto I_a \otimes \p_L (W_L ((0,\,F_\pm)))\otimes I_M\in \cU(\cH_f)
\]
is a strongly continuous unitary representation of $N$ on the Hilbert space $\cH_f$, acting locally on the net $\cC_\cI$.\\
ii) and iii). It is easy to check the DHR superselection criterion requirements for such automorphisms, relatively to the indicated nets. 
In particular, to show  the equivalence in iii) it is enough to observe that the map
\[
\r^I_F:=\ad \big(\p_f(W(F))\big)\mapsto \ad \big(\p_+(W(\psi_\infty (F)))\otimes \p_-(W(\psi_\infty (F)))\big)=:\r^\cO_F
\] 
defines an automorphism for the net $\widetilde{\cA}$ with the required properties. 
Positivity of the energy and Poincar\'e covariance of $\r^\cO_F$ is proved in \cite{sw70}. The M\"{o}bius covariance is proved using the chiral formulation and the representation of the M\"obius group given in equation (\ref{e:covariance reps}).
\qed
\brem\label{r:solisoli}
1) Notice that the time zero formulation of Proposition \ref{p:sectors autom} distinguishes between the different nature of the two families of DHR automorphisms obtained for the net $\cA$: the ones in i) are the restriction of solitonic automorphisms of the bigger net $\cC$; the ones in ii) instead are the restriction of DHR automorphisms of the bigger net $\cQ$. Similarly for the net $\widetilde{\cA}$ because of iii).
In the following Subsection \ref{ss:gaugess}, the net $\cA$ is characterized as the fixed-point subnet of $\cQ$ under the action of a compact gauge group $\cG_q$, i.e. $\cA= \cQ^{\cG_q}$. 
Using the discussion and terminology in  \cite{klx04}, we say that the automorphisms in i) give \emph{twisted} representations of $\cA$ and the ones in ii)  \emph{untwisted}, relatively to the representation of the net $\cQ$.
In contrast, the chiral approach describes the sectors as indistinguished restrictions of solitonic automorphisms on the light lines.  
\\
2)
In the general situation treated in \cite{mug99}, the split property for the net is required in order to construct \emph{disorder operators} that implement the solitonic automorphisms; in the Streater and Wilde case such operators are directly given as Weyl elements in the larger algebras.
\erem
\subsection{Gauge symmetry group}\label{ss:gaugess}
We consider in the sequel the gauge automorphisms defined on net $\cF_\cI$. 

\smallskip
The general theory in \cite{dhr69I,dhr69II}, the construction of the Weyl algebras sector automorphisms and the structure of simple current extensions  in ii) of Proposition \ref{p:split local lag}, suggest looking for gauge symmetries in the dual of the charge group $\dG$.
We also recall the results iii) and iv) in Proposition \ref{p: elem weyl rep}, where the properties of the elementary Weyl algebra on the symplectic space $L=C\oplus N \cong \bR^2_d$, in non-regular representation, are displayed. Of particular relevance is the Bohr compactification of $N$. 

\smallskip
Consider the introduced two Abelian groups: the charge group $\dG:=C\oplus Q\cong \bR^2_d$ furnished with the discrete topology and the additive group $\cG_0:=N\oplus R$ as defined in the symplectic isomorphism in Subsection \ref{ss:WSW model} and furnished with the usual topology of $\bR^2$. 

\smallskip
Consider moreover the Abelian group of \emph{all} characters on $\dG$, i.e. the group of all the functions $\c: \dG \to \bT$ such that $\c (s+s') = \c(s) \c(s')$ for $s,s'\in \dG$, with the identity function as neutral element and complex conjugation as inverse. If we furnish this group with the compact-open topology, that coincides with the pointwise convergence topology, $\dG$ being discrete, we obtain an Abelian compact group we denote by $\cG$, because of the Pontryagin duality theorem. 

Moreover, the group $\cG$ is isomorphic to the \emph{Bohr compactification of} $\cG_0$, i.e. the closure of the group $\cG_0$ in the compact-open topology defined above. 

Observe that the group $\cG_0$ is identified with the subgroup of the characters of $\dG$ that correspond to strongly continuous one dimensional representation of $\dG$, with the usual topology. This means that under the groups morphism
\beqn
\barr{rcl}
\c:&\cG_0 &\longrightarrow \cG\\
&(n, r)&\longmapsto \c_{(n, r)}\,, 
\earr
\eeqn
the elements of $\cG_0$ are the elements $\c_{(n, r)}$ of $\cG$ with the property that for $\l\in \bR$ the map $\c_{(n, r)}(\l(c,q))=e^{i\l(nc +rq)}, \, (c,q)\in \dG$ is continuous.
The properties collected in the following result, authorize to call $\cG$ \emph{the gauge group of the net $\cA$}, at least relatively to the net $\cF$. 
\bprop\label{p: two groups}
The map defined by
\beqn
\barr{rcl}
V:&\cG_0 &\longrightarrow \cU ( \cH_f )\\
&(n, r)&\longmapsto V(n, r):=I_a \otimes \p_L\big(W_L((0,\,n))\big)\otimes \p_M\big(W_M((r,\, 0))\big)\,
\earr
\eeqn
satisfies:
\bdes
\item[i)] $V$ is a strongly  continuous unitary representation of $\cG_0$ on $\cH_f$; 
\item[ii)] the invariant Hilbert subspace of $V(\cG_0)$ is $\cH_a$. Moreover, the representation $V$ leaves invariant $\O_f = \O_a \otimes\O_L\otimes \O_M$, the vacuum vector of $\cF_\cI$;
\item[iii)] on the net $\cF_\cI$ the adjoint action of $V$ implements an automorphism group  we indicate by the same symbol $\cG_0$, such that for $(n, r)\in \cG_0$ and $F \in V_f$ we have
\[
\a_{(n,r)} ( \p_f (W_f (F) )) = \ad  V (n,r) (\p_f (W_f (F)))= e^{-i (n F_c + r F_q)}\p_f (W_f (F))\,.
 \]
This automorphism group acts strictly locally on the net $\cF_\cI$ and on the 1+1-dimensional net $\widetilde{\cF}=\cF_+\,\timeinfty\, \cF_-$, commuting with the action of the M\"{o}bius group on the net $\widetilde{\cF}$, represented on $\cH_f \cong\cH_+\otimes\cH_-$ as in equation (\ref{e:mgroup f}) and acting on the subspace of fixed charges $(c_+,c_-)$ as in equation (\ref{e:covariance reps});
\edes
Moreover, for the Abelian compact group $\cG$ defined above, we have:
\bdes
\item[iv)] $\cG\cong \cG_c\times \cG_q$, where $\cG_c$ and $\cG_q$ are the Bohr compactifications of $N$ and $R$ respectively. The group $\cG$ is strong continuously represented on $\cH_f$, extending the representation $V$, and preserving the properties in ii) and iii) of the subgroup $\cG_0$;
\item[v)] the dual group of $\cG$ is $\dG$, the charge group;
\item[vi)] it holds $\cF_\cI^{\cG_q}=\cC_\cI$,\,\, $\cF_\cI^{\cG_c}=\cE_\cI$\, and\,  $\cF_\cI^{\cG} = \cB_\cI$.
\edes
\eprop
\prf
Strong continuity in i) and iv) holds because the group $\cG_0$ is represented by $V$ using only the regular subspace $N\oplus R\subset L\oplus M$ of the representation $\p_L\otimes\p_M$ and, by density, extends to elements in $\cG$ because of the universal property of Bohr compactification, see \cite[Chapter 16]{DI}.\\
ii) The invariance of the subspace $\cH_a$ and of the vector $\O_f$ hold because of the factorization property of the state $\o_f=\o_a\otimes\o_L\otimes\o_M$, and of its GNS representation, as illustrated in Proposition \ref{P:factor state weyl case}.\\
iii) For the elements in $\cG_0$ this is trivial, because the action is implemented by Weyl elements.  The commutation of the action with the action of the  M\"{o}bius group derives from the invariance of charges and the  $\bT$-valued action of the latter group.\\
iv) results from the discussion above. In particular, the locality of the action for elements of $\cG$ not in $\cG_0$ holds, because it is $\bT$-valued.\\
v) by the Pontryagin's duality theorem.\\
vi) follows from the general case in \cite{dhr69I}, averaging over the compact group $\cG$, or the indicated subgroups, with respect to the Haar measure, in the one dimensional representation fixed by specifying the charges of the Weyl generators.
\qed
\brem
Observe that the previous introduction of a non splitting isomorphism, i.e. of a regularizing element $T \in V_f$, has been also useful to simplify the proofs of the Propositions \ref{p:sectors autom} and \ref{p: two groups}. Actually this isomomorphism corresponds to a section $\dG \ni (c,q) \to \D_c / \inn \cA $, where $\D_c$ is the group of the covariant DHR automorphisms and $\inn \cA$ the one of the inner automorphisms of $\cA$. This section gives a group homomorphism, as said in \cite[Section IV]{dhr69II} and a similar choice is also made in a non trivial center situation, as illustrated in \cite{bl02}.
\erem
Having established the existence of the superselection automorphisms and their relative gauge groups, the following diagram concerns the full meaning of the simple current extension diagram seen in (\ref{e:SW general inclusion of net}):

\begin{footnotesize}
\beq\label{e:SW inclusion of net}
\barr{ccccc}
&&\cA\otimes \Zb&&\cB\rtimes \cU(C)\\
&&\Vert &&\Vert\\
 \cA=\cQ^{\cG_q}& \subset & \cB=\cE^{\cG_q}=\cC^{\cG_c}=\cF^\cG&\subset&\cC=\cF^{\cG_q}\\\\
 		\cap	&& \cap& & \cap \\\\
		\cQ&\subset&\cE=\cF^{\cG_c}& \subset&\cF\\
\Vert&&\Vert&&\Vert \\
\cA\rtimes \cU(Q)&&\cB\rtimes\cU(Q)&&\cB\rtimes\cU(Q)\rtimes\cU(C)
\earr
\eeq
\end{footnotesize}
Notice that    
\bitem
\item 
the vertical lines describe fixed-point restrictions under the action of the compact subgroup $\cG_q$; dually they account for (twisted) DHR sectors of $\cA_\cI$ and $\cB_\cI$ and  solitonic (automorphism) sectors for $\cC_\cI$;
\item 
the horizontal lines describe fixed-point restrictions under the action of the compact subgroup $\cG_c$; dually they account for (untwisted) DHR sectors of nets $\cA_\cI,\,\cB_\cI,\,\cQ_\cI$ and $\cE_\cI$.
\eitem
To conclude, it worth recalling that such a kind of diagrams is present in other situation where the sectors of a net are obtained along two different extension procedures, through \emph{partial field nets}, see e.g. the \emph{square of nets} in \cite{reh97,mug99}.  For a general explanation of these features in terms of \emph{braided crossed $G$-categories}, for $G$ a finite gauge group, see \cite{mug04}. Regarding the last cited paper, it is to observe that in the Streater and Wilde model only the discrete subgroup $N\cong \bR_d$ is present in the description of solitonic sectors, and not the entire gauge group. 

\bigskip
\noindent {\bf Acknowledgements.\,}
This paper is mainly based on my PhD thesis written under the supervision of John Roberts to whom I'm grateful for continuing interest and support. A partial list of people to whom I'm indebted for useful conversations and hints about the topics of this paper, includes Se\-ba\-stia\-no Carpi, Roberto Conti, Roberto Longo,  Michael M\"uger, Fernando Lled\'o, Gerardo Morsella, Gherardo Piacitelli, Giuseppe Ruzzi and Ezio Vasselli. 

%
%
%
%
%

{\footnotesize
\begin{thebibliography}{alpha}
%
\bibitem{ams93}
Acerbi, F., Morchio, G., Strocchi, F.: Infrared singular fields and nonregular representations of canonical commutation relation algebras.
J. Math. Phys. \textbf{34}, 899-914 (1993).

\bibitem{ams93a}
Acerbi, F., Morchio, G., Strocchi, F.: Theta Vacua, Charge Confinement and Charged Sectors from Nonregular Representations of CCR Algebras. Lett. Math. Phys. \textbf{27}, 1-11 (1993).

\bibitem{bg04}
Baumgaertel, H., Grundling, H.: Superselection in the presence of constraints. J. Math. Phys. \textbf{46}, no. 8, 34 pp. (2005).

\bibitem{bl02}
Baumgaertel, H., Lledo, F.: Dual group actions on C*-algebras and their description by Hilbert extensions. Math. Nachr. \textbf{239/240}, 11-27 (2002).

\bibitem{bl03}
Baumgaertel, H., Lledo, F.: Duality of compact groups and Hilbert C*-systems for C*-algebras with a nontrivial center. Internat. J. Math. \textbf{15}, 759-812 (2004).

\bibitem{BRI}
Bratteli, O., Robinson, D. W.: \emph{Operator Algebras and Quantum Statistical Mechanics. I} Springer, New York, 1987. Second Edition.

\bibitem{BRII}
Bratteli, O., Robinson, D. W.: \emph{Operator Algebras and Quantum Statistical Mechanics. II} Springer, New York, 1997. Second Edition.

\bibitem{bgl93}
Brunetti, R., Guido, D., Longo, R.: Modular Structure and Duality in Conformal Quantum Field Theory. Commun. Math. Phys.  \textbf{156}, 201-219 (1993).

\bibitem{bdlr92} 
Buchholz, D., Doplicher, S., Longo, R., Roberts, J.E.: A new look at Goldstone's theorem. Rev. Math. Phys. \textbf{Special Issue}, 49-83 (1992).

\bibitem{bmt88} 
Buchholz, D., Mack, G., Todorov, I.: The current algebra on the circle as a germ of local field theories. Nucl. Phys. B. (Proc. Suppl.) \textbf{5B}, 20-56 (1988).

\bibitem{car04}
Carpi, S.: On the representation theory of Virasoro Nets. Commun. Math. Phys.  \textbf{244}, 261-284 (2004).

\bibitem{cav98}
Cavallaro, S.: Multiplicity-free representations of commutative C*-algebras and spectral properties. ArXiv math.OA/9804090.

\bibitem{fabio03} 
Ciolli, F.: {\it Sui settori di Superselezione del Campo Libero Scalare a Massa nulla in 1+1 dimensioni}. Phd thesis, Universit\`a di Messina, 2003.

\bibitem{fabio4} 
Ciolli, F.: Massless scalar free Field in 1+1 dimensions II: Net Cohomology and Completeness of Superselection Sectors. arXiv:0811.4673.


\bibitem{dm04}
Derezi{n}ski, J., Meissner K.A.: Quantum massless field in 1+1 dimensions. Lecture Notes in Phys., \textbf{690}, 107-127. Springer, Berlin, 2006. 

\bibitem{DI}
Dixmier, J.: \emph{Les C*-alg\`ebres et leurs repr\'esentations.} Gautier Villars, Paris, 1964. 

\bibitem{DII}
Dixmier, J.: \emph{C*-algebras}. Amsterdam : North Holland, 1977.

\bibitem{dhr69I}
Doplicher, S., Haag, R., Roberts, J. E.: Fields, observables and gauge
transformation I. Commun. math. Phys. \textbf{13}, 1-23 (1969).

\bibitem{dhr69II}
Doplicher, S., Haag, R., Roberts, J. E.: Fields, observables and gauge transformation II. Commun. math. Phys. \textbf{15}, 173-200 (1969).

\bibitem{dhr71}
Doplicher, S., Haag, R., Roberts, J. E.: Local observables and particle statistics I. Commun. math. Phys. \textbf{23}, 199-230 (1971).

\bibitem{DL}
Doplicher, S. (Editor), Longo, R. (Editor): \emph{Noncommutative Geometry.} Martina Franca, Italy 2000. Lecture Notes in Mathematics \textbf{1831}. Springer-Verlag, 2004.

\bibitem{dr90}
Doplicher, R., Roberts, J. E.: Why there is a field algebra with a compact gauge group describing the superselection structure in particle physics. Commun. math. Phys. \textbf{131}, 51-107 (1990).


\bibitem{EK}
Evans, D.E., Kawahigashi, Y.: {\em Quantum symmetries on operator algebras.}  Oxford: Oxford University Press, 1998. 

\bibitem{fro76}
Fr\"ohlich, J.: New super-selection sectors (`Soliton-states') in two-dimensional Bose quantum field models. Comm. Math.Phys. \textbf{47}, 269-310 (1976).

\bibitem{gru97}
Grundling, H.: A Group Algebra for Inductive Limit Groups. Continuity Problems of the Canonical Commutation Relations. Acta Appl. Math. \textbf{42}, 107-145 (1997).

\bibitem{H}
Haag, R.: \emph{Local Quantum Physics.} Texts and Monographs in Physics, Berlin: Springer, 1992, 1996. 

\bibitem{hal04}
Halvorson, H.: Complementarity of representations in quantum mechanics. Studies in History and Philosophy of Modern Physics \textbf{35}, 45-56 (2004).

\bibitem{her98}
Herdegen, A.: Semidirect product of CCR and CAR algebras and asymptotic states in quantum electrodynamics. J. Math. Phys. \textbf{39}, 1788-1817 (1998).

\bibitem{hl82}
Hislop, P. D., Longo, R.: Modular Structure of the Local Algebras Associated with the Free Massless Scalar Field Theory. Commun. Math. Phys. \textbf{84}, 71-85 (1982).

\bibitem{KRI}
Kadison, R. V., Ringrose, J. R.: \emph{Fundamentals of the Theory of Operator Algebras. I}. Graduate Studies in Mathematics. American Mathematical Society, Providence, 1997.

\bibitem{KRII}
Kadison, R. V., Ringrose, J. R.: \emph{Fundamentals of the Theory of Operator Algebras. II}. Graduate Studies in Mathematics. American Mathematical Society, Providence, 1997.

\bibitem{kl04}
Kawahigashi, Y., Longo, R.: Classification of Two-dimensional Local Conformal Nets with $c<1$ and 2-cohomology Vanishing for Tensor Categories. Commun. Math. Phys. \textbf{244}, 63-97 (2004).

\bibitem{kl04a}
Kawahigashi, Y., Longo, R.: Local conformal nets arising from framed vertex operator algebras.  Adv. Math.  \textbf{206}, 729-751 (2006).


\bibitem{klm01}
Kawahigashi, Y., Longo, R., M\"uger, M.: Multi-interval Subfactors and Modularity of Representations in Conformal Field Theory. Commun. Math. Phys. \textbf{219}, 631-669 (2001).

\bibitem{klx04}
Kac, V. G., Longo, R., Xu, F.: Solitons in Affine and Permutation Orbifolds. Commun. Math. Phys. \textbf{253}, 723-764 (2004).


\bibitem{L}
Longo, R. (ed.): \emph{Mathematical Physics in Mathematics and Physics. Quantum and Operator Algebraic Aspects}. Fields Institute Communications, Volume \textbf{20}. American Mathematical Society, 2001.

\bibitem{lx04}
Longo, R., Xu, F.: Topological Sectors and a Dichotomy in Conformal Field Theory. Commun. Math. Phys. \textbf{251}, 321-364 (2004).


\bibitem{mstv}
Manuceau, J., Sirugue, M., Testard, D., Verbeure, A.: The smallest C*-algebra \ for\ Canonical\ Commutation \ Relations. Com\-mun. Math. Phys. \textbf{32}, 231-243 (1973).

\bibitem{mug98}
M\"uger, M.: Superselection Structure of Massive Quantum Field Theories in (1+1)-dimensions. Rev. Math. Phys. {\bf 10}, 1147-1170 (1998).

\bibitem{mug99}
M\"uger, M.: On Soliton Automorphisms in Massive and Conformal Theories. Rev. Math. Phys. \textbf{11}, 337-359 (1999). 

\bibitem{mug01}
M\"uger, M.: Conformal Field Theory and Doplicher-Roberts Reconstruction. In: \cite{L}. 

\bibitem{mug04}
M\"uger, M.: Conformal Orbifold Theories and Braided Crossed G-Categories.  Comm. Math. Phys.  \textbf{260},  727-762 (2005).

\bibitem{reh97} 
Rehren, K.-H.: Spin-statistics and CPT for solitons. Lett. Math. Phys. \textbf{46},  95-110, (1998).

\bibitem{reh00} 
Rehren, K.-H.: Chiral Observables and Modular Invariants. Commun. Math. Phys. \textbf{208},  689-712, (2000).

\bibitem{rob76}
Roberts, J. E.: Local cohomology and superselection rules. Comm. Math. Phys. \textbf{51}, 107-119 (1976).

\bibitem{rob77}
Roberts, J. E.: Mathematical Aspects of Local Cohomology. Proceedings of the Colloquium on Operator Algebras and their Applications to Mathematical Physics, 321-332, CNRS, Marseille, 1977.

\bibitem{rob98}
Roberts, J. E.: Lecture notes of 1997/98 course held at Dipartimento di Matematica,  Universit\`a di Roma 2 Tor Vergata, Roma (1998).

\bibitem{rob00}
Roberts, J.E.: More lecture on Algebraic Quantum Field Theory. In \cite{DL}.

\bibitem{sch06}
Schroer, B.: Two-dimensional models as testing ground for principles and concepts of local quantum physics.  Ann. Physics  \textbf{321}, 435-479 (2006). 

\bibitem{sla72}
Slawny, J.: On factor representations and the C*-algebra of Canonical Commutation Relations. Comm. Math. Phys. \textbf{24}, 151-170 (1972).

\bibitem{sw70}
Streater, R. F., Wilde, I. F.: Fermion states of a Boson field. Nucl. Phys. \textbf{B24}, 561-575 (1970).

\bibitem{TI}
Takesaki, M.: \emph{Theory of Operator Algebras I}. Springer, 2001.
\bibitem{TII}
Takesaki, M.: \emph{Theory of Operator Algebras II}. Springer, 2002.
\bibitem{TIII}
Takesaki, M.: \emph{Theory of Operator Algebras III}. Springer, 2002.

\bibitem{tn92}
Thirring, W., Narnhofer, N.: Covariant QED without indefinite metric. Rev. Math. Phys. \textbf{Special Issue}, 197-211 (1992).
%
\end{thebibliography}
}
\end{document}